\documentclass[reprint,superscriptaddress,nofootinbib,amsmath,amssymb,aps,prd]{revtex4-2}

\usepackage{bm, graphics, graphicx, epsfig, soul, latexsym, hyperref ,multirow}
\usepackage{comment}
\usepackage{subfigure}
\usepackage{graphicx,color}
\usepackage{dcolumn}
\usepackage{bm,amsmath, amssymb,}
\usepackage{mathrsfs}
\usepackage{bigints}
\usepackage{float}
\usepackage{multirow}
\usepackage{placeins}
\usepackage{subfigure}
\usepackage[normalem]{ulem}
\usepackage{booktabs}
\usepackage{tabularx}

\usepackage{amsmath}

\newcommand{\be}{\begin{equation}}
\newcommand{\ee}{\end{equation}}

\begin{document}

\title{Probing black hole `charge' from the binary black hole inspiral}

\author{N. V. Krishnendu}
\email{krishnendu.nv@icts.res.in}
\affiliation{International Centre for Theoretical Sciences (ICTS), Survey No. 151, Shivakote, Hesaraghatta, Uttarahalli, Bengaluru, 560089, India}
\author{Sumanta Chakraborty}
\email{tpsc@iacs.res.in}
\affiliation{School of Physical Sciences, Indian Association for the Cultivation of Science, Kolkata 700032, India}

\date{\today}

\begin{abstract}
Recent gravitational wave (GW) observations have enabled us to look beyond the standard paradigm of gravitational physics, namely general relativity (GR). Along with the mass and the angular momentum, which typical astrophysical black holes (BHs) are endowed with, theories beyond GR generically induce `charge' to these BHs. Notably, for BHs carrying the extra `charge' hair, we expect the BH absorption effects to modify accordingly and alter the tidal heating terms. Hence, the inclusion of the corrections in the GW waveform model, arising from the BH `charge', allows us to test the consistency of the observed binaries with Kerr BHs in GR. We compute the explicit dependence of the binary inspiral phase on the `charge' parameter arising from the tidal heating effect and study the measurability of the same from GW observations of binary mergers. Specifically, we employ the {\tt TaylorF2} waveform model, which accurately models the inspiral evolution of an aligned-spin binary merger, and Bayesian analysis-based GW data inference to measure the `charge' parameter for a selected set of detected binaries. We also present a detailed simulation study to investigate the possibility of measuring the charge parameter from binaries with different masses, spins and source locations. The analysis of selected GW events from the third GW transient catalogue shows that the `charge' parameter constraints are poor from the observed signals with the current sensitivity. In contrast, the simulation studies indicate that the spinning binaries with significant mass asymmetry provide the best constraints on the BH `charge' parameter. Finally, we study the prospects of measuring the BH `charge' parameter from a future GW detector with improved sensitivity. 
\end{abstract}
\maketitle
\section{Introduction}\label{intro}
After finishing the three successful observing runs, the ground-based gravitational wave (GW) detectors confirmed the detections of more than ninety compact binary coalescing events~\cite{AdvLIGO, AdvVirgo, virgo,ligo, GWTC3-catalog, GWTC-2-catalog, GWTC-2.1-catalog,4OGC, IAS-3}. Then, the first part of the fourth observation run concluded by reporting around eighty GW events and details are available in the public alert system~\cite{RICHABBOTT2021100658}. From the detection and further dedicated analysis, it is evident that the signals consist of three types of merger events --- (a) binary black holes (BHs), (b) binary neutron stars (NSs), and (c) BH-NS systems~\cite{GW170817, GW190425, GW190521, NS_BH_GW200105_GW200115}. Even though the observed signals are consistent with these compact objects being described by either BHs or NSs within the framework of general relativity (GR) and show no clear evidence of the presence of physics beyond the standard paradigm~\cite{GWTC-1-TGR, GWTC-2-TGR, GWTC-3-TGR, TOGGW150914}, however, given the statistical uncertainties in these measurements, there is still ample room for alternate theories of gravity and exotic compact objects~\cite{Berti_2015, Cardoso:2017cfl}. 

Despite the monumental success of GR in all the observational tests so far~\cite{EventHorizonTelescope2020TGR, TOGGW150914}, the prime reasons to look for alternatives of GR are threefold --- (a) Einstein's equations predict singularities, where the theory itself breaks down, (b) solutions of Einstein's equations involve Cauchy horizon, leading to uncertain future for the classical theory, (c) consistency with observations require postulating the existence of yet undetected dark matter and exotic dark energy. Besides, there are other important issues related to the energy scale at which gravity dominates, as this scale is much above the energy scales associated with the other fundamental interactions. This leads to the well-known hierarchy problem, where the stabilization of the Higgs mass requires an exorbitant fine-tuning of the order of one part in $10^{15}$. The models involving extra spatial dimensions are among several alternatives to get around this fine-tuning problem. In this work, we explore the implications of the existence of an extra spatial dimension on the inspiral regime of the binary BH merger events detected by the LIGO-Virgo-KAGRA collaboration. The presence of this extra spatial dimension modifies the gravitational field equations on the four-dimensional spacetime (known as the \emph{brane}), as the projection of the five-dimensional (known as the \emph{bulk}) Einstein's equations inherit additional contributions from the bulk. Consequently, the BHs on the brane are characterized by an additional hair, other than the mass and the spin. This hair is identical in appearance to that of the Maxwell charge but differs by an overall sign, which becomes a distinctive signature of the existence of extra spatial dimensions~\cite{Aliev:2005bi, Harko:2004ui, Dadhich:2000am, Shiromizu:1999wj}. It is worth pointing out that a similar contribution, albeit with an opposite sign, appears in the context of Einstein-Maxwell theories, scalar coupled Maxwell theories~\cite{Babichev:2013cya, Barrientos:2017utp, Babichev:2015rva, Maeda:2006hj} and also in $f(T)$ theories of gravity \cite{Faraoni:2010pgm, Capozziello:2012zj}. The presence of the extra `charge' in the BH solution can induce differences in the evolution of a compact binary system, making their GW signatures utterly different from that of the Kerr BH. Specifically, by focusing on the inspiral dynamics of the binary, we study the significance of the `charge' hair on the tidal heating effects present in the GW waveform measured at infinity. 

{\it Tidal heating} is an absorption effect, which arises due to the absorption of external GWs by compact objects in a binary, leading to time evolution of the mass and angular momentum of the compact objects, leaving observational imprints~\cite{Hughes:1999bq, Hartle1973, Isoyama:2017tbp, Chatziioannou:2012gq, Datta:2019epe, Datta:2020rvo, Datta:2019euh, Alvi:2001mx, Poisson:2004cw, Saketh:2022xjb, Datta:2023wsn, Mukherjee:2023pge, Datta:2020gem, Maselli:2017cmm}. The existence of the `charge' hair will modify the absorption of GWs by the BHs on the brane, which will reflect in the orbital evolution of the binary in the inspiral phase. Following this, we propose a novel test to measure the `charge' parameter from the observed inspiral evolution of the binary coalescence events and demonstrate the measurability over a set of simulated binary signals. The inspiral measurements of the BH `charge' provide a unique test of the nature of compact binary because for astrophysical BHs, the net electric charge is expected to be zero, or even if there is a small electric charge present, that gets shielded away quickly~\cite{Feng:2022evy, Pina:2022dye, Bozzola:2020mjx}. Therefore, the existence of a `charge' hair, along with its overall sign, is a characteristic of the presence of an extra spatial dimension. 

There have been several previous attempts in the literature to look for the existence of such a `charge' in various observations involving BHs. These include --- (a) the shadow of M87* and SgrA* as observed by the Event Horizon Telescope collaboration~\cite{Neves2020M87,Zajacek2018SgrAReview,Zakharov2021EHT2017}, (b) the X-ray luminosity from the accretion disc of quasars~\cite{Maselli:2014fca,Stuchlik:2008fy, Banerjee:2017hzw,Banerjee:2021aln}, and (c) the GW observations from binary coalescence~\cite{Barausse2015wia,Andriot2017,Chakraborty2017,Chakravarti2019,Gu:2023eaa,Gupta:2021rod,Carullo:2021oxn}. Intriguingly, the observations involving BH shadow and accretion disc around quasars mildly favour the existence of a negative `charge' compared to the Kerr scenario in GR. The ringdown signal, on the other hand, provides little information, and both GR and extra dimensions remain viable alternatives. In Ref.~\cite{MishraAkash_I}, the BH ringdown waveform has been obtained by numerically solving the perturbation equations for the braneworld BH and determining the associated quasi-normal modes~\cite{MishraAkash_I} with the explicit BH `charge' dependence. Subsequently, this analysis has been extended in~\cite{MishraAkash_II}, and constraints on the `charge' parameter have been derived from the events observed through the first three observing runs of advanced LIGO and Virgo detectors. In a different perspective,~\cite{UddeeptaDeka2024} demonstrated yet another method to constrain the `charge' parameter by looking into the GW micro-lensing signature of the charged lens. It is also possible through anti-de Sitter/conformal field theory correspondence (AdS/CFT) correspondence to argue that horizons for BHs on the brane must be reflective \cite{Dey:2020lhq,Dey:2020pth}, however, in this work, we will stick to the classical BH picture, returning to the reflective nature of horizon in future work. To summarise, we focus on the inspiral dynamics of the binary and study the distinctive features in the BH absorption spectra due to the presence of the `charge' parameter arising from the existence of an extra spatial dimension considering stellar mass binary BH mergers.

The detection and parameter inference necessitate employing waveform models based on GR that accurately predict the binary evolution, including various physical effects of binary BHs in GR. One must validate these GW waveform models to regimes inaccessible to GR and its predictions to look for beyond-GR effects in the GW data. However, the theory-agnostic tests of GR are routinely employed to check the consistency of GR with the observed data where generic parametric deviations are introduced in the GW waveform model without assuming any specific modified gravity models~\cite{TOGGW150914, GWTC-1-TGR, GWTC-2-TGR, GWTC-3-TGR}. Suppose GR is the correct theory of gravity; in this case, the measurement uncertainties on the parametric deviation coefficients will be consistent with GR prediction, and any departure will lead to further elaborate analyses. In the ideal scenario, one may start with an alternate gravity model, compute the GW waveform model for a merging binary in that particular theory, and use it for the analysis. Despite multiple efforts, such a complete model has yet to be available. In the middle ground, one can combine the orbital evolution information from GR, calculate additional contributions from beyond GR effects and estimate the GW waveform model. Analysis employing such a modified waveform model will provide a consistency check with the observed data and GR predictions. In this study, we focus on such a scenario. 

To do so, we calculate the BH absorption effects of a braneworld BH and modify the post-Newtonian inspiral phase by appropriately adding the tidal charge contributions. We keep the `charge' parameter as a free parameter and measure it from the data. Before discussing the real GW analysis, we demonstrate the method by simulating a set of binaries with various masses, spins and locations. The simulations indicate that it is possible to measure the `charge' parameter for spinning binaries with mass asymmetry if we consider current-ground-based GW detectors with their plus-era (O5) sensitivity~\cite{KAGRA:2013rdx}. Further, we describe the method's applicability for inspiral-dominated GW events detected through the first three observing runs of LIGO-Virgo detectors. For completeness we will discuss both the scenarios involving positive as well as negative `charged' hairs. 

The paper is organized as follows: we give a brief outline of our geometrical setup in Sec.~\ref{subsec:cargedBH}, and then discuss the waveform model, a short description of the Bayesian analysis and a note on the details of binary simulations in Sec.~\ref{analysis}. The results from simulated binary signals are presented in Sec.~\ref{res1}, while our bounds from the GW signals of the detected binary merger events have been presented in Sec.~\ref{res2}. Finally, we conclude by summarising our findings and listing future plans in Sec.~\ref{sec:summary}. 
\section{Background geometry with charge: Implications for tidal heating}\label{subsec:cargedBH}
In this section, we present the background geometry of a braneworld BH, with the charge term, and also discuss the implications of this on the tidal heating phenomenon. In the braneworld scenario, the higher dimensional spacetime (\emph{bulk}) satisfies Einstein's equations, while the gravitational field equations on the embedded four-dimensional hypersurface (\emph{brane}), on which the standard model fields live become,
\begin{align}
~^{(4)}G_{\mu \nu}+E_{\mu \nu}=0~.
\end{align}
Here, $~^{(4)}G_{\mu \nu}$ is the four-dimensional Einstein tensor on the brane hypersurface, $E_{\mu \nu}\equiv~^{(5)}W_{ABCD}e^{A}_{\mu}n^{B}e^{C}_{\nu}n^{D}$ is the projection of the five-dimensional Weyl tensor $~^{(5)}W_{ABCD}$ on the brane hypersurface, where $e^{A}_{\mu}$ are the projectors and $n^{B}$ is the normal vector to the brane hypersurface. Owing to the symmetries of the Weyl tensor, it follows that $E^{\mu}_{\mu}=0$, and Bianchi identity demands $\nabla_{\mu}E^{\mu}_{\nu}=0$. Both of these properties are akin to the energy-momentum tensor of the electromagnetic field, except for an overall sign. This is because the energy-momentum tensor sits on the right-hand side of Einstein's equation, acting as the source of gravity, while the Weyl tensor $E_{\mu \nu}$ sits on the left-hand side, which mimics a source with energy-momentum tensor $-E_{\mu \nu}$. Thus, braneworld BH depicts vacuum spacetime but resembles Kerr-Newman spacetime with an overall negative sign in front of the charge term. Therefore, the spacetime geometry of a rotating braneworld BH takes the form, 
\begin{align}
ds^{2}&=-\frac{\Delta}{\Sigma}(dt- a \sin^2\theta\, d\phi)^2+\,\Sigma\left[\frac{dr^2}{\Delta}+\,d\theta^2\right] 
\nonumber
\\
&\qquad +\frac{\sin^2\theta}{\Sigma}\left[a\,dt-(r^2+a^2)d\phi\right]^2~.
\end{align}
The above metric depicts a rotating BH with mass $M$, angular momentum $J=aM$ and braneworld `charge' $\mathcal{Q}^{\rm BH}$. The metric functions appearing in the above line element involve two unknown functions $\Delta$ and $\Sigma$, defined as $\Delta\equiv r^{2}+a^{2}-2Mr-\mathcal{Q}^{\rm BH}$ and $\Sigma\equiv r^{2}+a^{2}\cos^{2}\theta$. Note that, for the case of Kerr-Newman BH, the parameter $\mathcal{Q}^{\rm BH}$ can be identified with the negative of the square of the electric charge $Q$ of the BH, such that $\mathcal{Q}^{\rm BH}|_{\rm KN}=-Q^{2}$. The rotating braneworld BH inherits two horizons, located at $r_{\pm}=M\pm \sqrt{M^{2}-a^{2}+\mathcal{Q}^{\rm BH}}$ obtained by solving the equation $\Delta=0$. Intriguingly, even if $a>M$, for non-zero values of $\mathcal{Q}^{\rm BH}$, the outer horizon $r_{+}$ exists, in stark contrast to that of the Kerr-Newman BH.   

Gravitational perturbation of the background geometry, depicting a rotating BH spacetime on the brane, can be described by Newman-Penrose scalars $\Psi_{0}$ and $\Psi_{4}$. Since our interest is in the physics of the horizon, namely in determining the GW flux going down the horizon, we will work with the Newman-Penrose scalar $\Psi_{0}$. In general, for the Kerr-Newman-like spacetimes, the angular and the radial parts of the Newman-Penrose scalars cannot be separated. However, in the present context, perturbation of the `source' term $E_{\mu \nu}$ is directly proportional to the ratio of the bulk and the brane curvature length scales, which can be ignored for all practical purposes. Therefore, the Weyl scalar $\Psi_{0}$, describing gravitational perturbation of a rotating braneworld BH, can be expressed as,
\begin{align}
\Psi_{0}=\int d\omega\sum_{\ell=0}^{\infty}\sum_{m=-\ell}^{\ell}~_{2}S_{\ell m}(\theta)R_{\ell m}(r)e^{-i\omega t}e^{im\phi}~,
\end{align}
where the angular part $_{2}S_{\ell m}(\theta)$ satisfies the following differential equation, 
\begin{align}
\frac{1}{\sin \theta}&\dfrac{d}{d\theta}\left[\sin \theta\dfrac{d  {}_{2}S_{\ell m}}{d\theta}\right]+ \Big[(a\omega \cos\theta)^{2}- 4a\omega \cos\theta+2
\nonumber
\\
&\qquad +{}_{2}A_{\ell m} -\dfrac{(m+2\cos \theta)^{2}}{1-\cos^{2}\theta}\Big]  {}_{2}S_{\ell m}=0~,
\end{align}
which coincides with the equation satisfied by the spin-weighted spherical harmonics. Here $_{2}A_{\ell m}$ is the separation constant between the radial and the angular parts. The radial function $R_{\ell m}(r)$, on the other hand, satisfies the following differential equation \cite{Teukolsky1972}. 
\begin{align}
\frac{1}{\Delta^{2}}&\dfrac{d}{dr}\left[\Delta^{3}\dfrac{dR_{\ell m}}{dr}\right]+\Big[\dfrac{K^{2}-4i(r-M)K}{\Delta}
\nonumber
\\
&\qquad \qquad \qquad \qquad + 4i\dfrac{dK}{dr}-\lambda \Big]R_{\ell m} =0~,
\end{align}
where, $K\equiv (r^{2}+a^{2})\omega-am$ and $\lambda= {}_{2}A_{\ell m}+(a\omega)^{2}-2am\omega$, is related to the separation constant $_{2}A_{\ell m}$ appearing in the angular part $_{2}S_{\ell m}$. 

Consider now a binary system involving two braneworld BHs, characterized by masses $M_{1}$ and $M_{2}$, angular momentum $J_{1}$ and $J_{2}$, as well as `charge' parameter $\mathcal{Q}^{\rm BH}$. In the context of braneworld BH, the charge $\mathcal{Q}^{\rm BH}$ depends on the length of the extra dimension, and hence is an invariant quantity for all BHs. While for positive values of $\mathcal{Q}^{\rm BH}$, we assume identical values of the charge for both the BHs in the binary, possibly due to some overall equilibrium. During the inspiral of this binary system around one another, each of these BHs will absorb a part of the emitted GW radiation in the centre of the mass frame, leading to a rate of change of mass $M$, angular momentum $J$, and area $A$. This is known as tidal heating. In order to determine the above rate of changes of BH parameters for the first BH, we must solve the radial equation near the horizon $r_{+1}$ and impose purely ingoing boundary conditions at the horizon. This fixes one arbitrary constant appearing in the solution of the Teukolsky equation. In order to fix the other, we need to work in a regime where $M_{1}\ll r \ll b$, with $b$ being a typical distance between the binary BHs, and one imposes the following boundary condition,
\begin{align}
\Psi_{0}=8\pi \frac{\sqrt{6}M_{2}}{5b^{3}}\sum_{m=-2}^{2}\,_{2}Y_{2m}(\theta,\phi)\,_{2}Y^{*}_{2m}(\theta_{0},\phi_{0})~.
\end{align}
To determine the tidal heating associated with the second BH, we simply have to interchange $M_{1}\leftrightarrow M_{2}$, and $J_{1}\leftrightarrow J_{2}$, respectively. With the above boundary condition, one can uniquely solve for the Weyl scalar $\Psi_{0}$ and transform the same to the Hartle-Hawking frame, thereby determining the rate of change of area in terms of $|\Psi_{0}|^{2}$. The corresponding rate of change of mass can be derived using the laws of BH mechanics, which relates the rate of change of mass and angular momentum to the rate of change of area, yielding \cite{Chakraborty:2021gdf}, 
\begin{align}
\frac{dM_{1}}{dt}&=\left(\dfrac{dE}{dt}\right)_{\rm N}\left(\frac{M_{1}}{M}\right)^{3}\frac{v^{5}}{4}
\Bigg\{-\chi_{1}\left(\hat{\bf{L}}_{\rm orb}\cdot \hat{\bf{J}}_{1}\right)
\nonumber
\\
&\qquad +2\frac{v^{3}}{M}\left(r_{+1}+\frac{\mathcal{Q}^{\rm BH}}{2M_{1}}\right)\Bigg\}\Bigg[1+3\chi_{1}^{2}
\nonumber
\\
&\qquad \qquad +\frac{\mathcal{Q}^{\rm BH}}{M_{1}^{2}}\left(2+3\chi_{1}^{2}+\frac{\mathcal{Q}^{\rm BH}}{M_{1}^{2}}\right)\Bigg]~,
\end{align}
where $(dE/dt)_{\rm N}=(32/5)\eta^{2}v^{10}$ is the energy loss due to GWs arising from the quadrupole approximation. Further, we have defined $M\equiv M_{1}+M_{2}$ and $\eta\equiv(M_{1}M_{2}/M^{2})$, while the relative velocity of the binary BH system is given by $v=\sqrt{M/b}$. A similar expression can be derived for $(dM_{2}/dt)$, by simply the $M_{1}\leftrightarrow M_{2}$ exchange. Thus, the rate of change of mass, in comparison to the quadrupolar rate of change of energy, depends on post-Newtonian (PN) terms of two distinct orders, at $2.5$ PN (or, equivalently $v^{5}$), 
\begin{align}\label{25pN}
A_{i}^{(5)}&\equiv \left(\frac{M_{i}}{M}\right)^{3}\chi_{i}\left(\hat{\bf{L}}_{\rm orb}\cdot \hat{\bf{J}}_{i}\right)\Bigg[1+3\chi_{i}^{2}
\nonumber
\\
&+q^{\rm BH}\left(\frac{M^{2}}{M_{i}^{2}}\right)\left\{2+3\chi_{i}^{2}+q^{\rm BH}\left(\frac{M^{2}}{M_{i}^{2}}\right)\right\}\Bigg]~,
\end{align}
and at $4$ PN (or, equivalently $v^{8}$), 
\begin{align}\label{5pN}
A_{i}^{(8)}&\equiv \left(\frac{M_{i}}{M}\right)^{4}\Bigg[1+\sqrt{1-\chi_{i}^{2}+q^{\rm BH}\left(\frac{M^{2}}{M_{i}^{2}}\right)}
\nonumber
\\
&\qquad +\frac{q^{\rm BH}}{2}\left(\frac{M^{2}}{M_{i}^{2}}\right)\Bigg]
\Bigg[1+3\chi_{i}^{2}+q^{\rm BH}\left(\frac{M^{2}}{M_{i}^{2}}\right)
\nonumber
\\
&\qquad \times \left\{2+3\chi_{i}^{2}+q^{\rm BH}\left(\frac{M^{2}}{M_{i}^{2}}\right)\right\}\Bigg]~.
\end{align}
Here we have defined, $q^{\rm BH}\equiv (\mathcal{Q}^{\rm BH}/M^{2})$, $\chi_{i}=a_{i}/M_{i}$ and $i=\{1,2\}$. Thus the rate of change of mass for the $i$th component of the binary BH system can be expressed as,
\begin{align}
\frac{dM_{i}}{dt}&=\left(\dfrac{dE}{dt}\right)_{\rm N}\left[-A_{i}^{(5)}\frac{v^{5}}{4}+A_{i}^{(8)}\frac{v^{8}}{2}\right]~.
\end{align}
Therefore, the total flux going into the horizon of the braneworld BH becomes, 
\begin{align}\label{flux_horizon}
F_{\rm BH}&=\sum_{i}\left(\frac{dM_{i}}{dt}\right)
=\left(\dfrac{dE}{dt}\right)_{\rm N}\left[-\Psi_{5}\frac{v^{5}}{4}+\Psi_{8}\frac{v^{8}}{2}\right]~.
\end{align}
Here, we have defined, $\Psi_{5}\equiv \sum_{i}A_{i}^{(5)}$ and $\Psi_{8}\equiv \sum_{i}A_{i}^{8}$, for notational convenience. As we will demonstrate, these two quantities will be central to the phase evolution in the presence of tidal heating. 

At this outset, let us discuss previous bounds on the tidal charge parameter $q^{\rm BH}$ and the implications of such bounds on the size of the extra dimension~\cite{Chamblin:2000ra}. In the context of GW observations, based on the ringdown part of the signal, \cite{Carullo:2021oxn, MishraAkash_II, MishraAkash_I} provides bounds on the tidal charge parameter. The $90\%$ confidence contours, for the majority of GW observations, extended beyond $q^{\rm BH}=\pm 0.5$. Besides, the lensing of GWs can also constrain the tidal charge $q^{\rm BH}$, but again for the braneworld models the constraints are weak $q^{\rm BH}\leq -0.9$, while for electromagnetic theories the constraints are better $q^{\rm BH}\leq 0.5$~\cite{UddeeptaDeka2024}. From electromagnetic observations as well, e.g., the measurement of BH shadow constrains the tidal charge as $q^{\rm BH}=-0.1^{+0.6}_{-0.5}$ \cite{Banerjee:2022jog, Banerjee:2019nnj}. These suggest that the constraints on the charge parameter, irrespective of its origin, are weak and are the prime motivation to choose the prior within the range $q^{\rm BH}\in (-1,1)$. Further note that, in the braneworld scenario, the brane is obtained by embedding the four-dimensional spacetime within a five-dimensional bulk. Therefore, the charge $q^{\rm BH}$ gets naturally connected to $\chi$, the size of the extra dimension. For example, the following bound: $q^{\rm BH}\leq -0.5$, translates into $(\chi/\ell) \lesssim 0.63$ \cite{Chamblin:2000ra}, where $\ell$ depicts the ratio of the five-dimensional and the four-dimensional gravitational constants.

Having determined the effect of the charge term in the spacetime metric, arising from extra spatial dimension, on the perturbation equation governing the gravitational perturbation of the brane. The corresponding fluxes through the horizons of the braneworld BHs orbiting each other get corrected at 2.5 PN and 4 PN levels due to the charge term. Given the above modification to the horizon flux, we wish to determine the corresponding modifications to the GW phase in the next section.   
\section{Details of the waveform model and parameter estimation}
\label{analysis}
In this section, we will present detailed discussion regarding the waveform modeling for braneworld BH, with special emphasis on the effect of charge inherited from extra spatial dimension. Then we  proceed to the parameter estimation details. 
\subsection{Waveform model}\label{subsec:waveform}

The GW waveform model from a coalescing compact binary signal in the frequency domain can be schematically represented as,
\begin{equation}
\tilde{h}(f)= \mathcal{C}\mathcal{A}(f) e^{i\{\psi_{\rm test}(f)+ \,\delta\psi(f)\}}~,
\label{eq:wf}
\end{equation}
where $\mathcal{C}$ is an overall constant, $\mathcal{A}(f)$ is the amplitude of the GW, $\psi_{\rm test}(f)$ is the phase of the GW in the test particle approximation~\cite{ABFO08,BIOPS2009}, and $\delta \psi(f)$ is the contribution to the phase due to finite size effects of the binary BH system. In the inspiral regime, the amplitude of the GW follows from the relation $\mathcal{A}(f)\sim D_{\rm L}^{-1} M_{c}^{5/6}f^{-7/6}$ at leading order, where $D_{\rm L}$ is the luminosity distance between the observer and the source of the GW, with $M_{c}=(M_{1}M_{2})^{3/5}(M_1+M_2)^{-1/5}$ being the chirp mass of the binary. The frequency-independent factor $\mathcal{C}$ carries information about the source location and orientation of the source with respect to the detector, through the antenna pattern functions. For a compact binary signal, the GW phase plays a crucial role in detecting and analysing the signal. Hence, it is important to model them with the maximum available accuracy. In our case, the first term, ${\psi_{\rm test}(f)}$, accounts for the `point-particle' contributions, whereas ${\delta\psi}(f)$ represents the extra phase contributions that arise due to tidal heating, or, the BH absorption effect. Among these terms, ${\psi_{\rm test}(f)}$ is taken to be accurate upto 3.5 PN~\cite{Mishra:2016whh}, and ${\delta\psi}(f)$ has contributions at 2.5 PN, 3.5 PN and 4 PN orders. Altogether, these phase contributions accurately model the binary dynamics to the respective PN orders of aligned spin braneworld BHs in a binary system. 

The explicit expression for the phase $\delta \psi(f)$, due to tidal heating, can be obtained using the phase formula of \cite{Datta:2020rvo,Samanwaya2022}, and then using the flux through the horizon due to tidal heating, derived in Eq. \eqref{flux_horizon}. The final expression reads (for a detailed derivation, see Appendix \ref{app_Phase_tidal_heating}), 
\begin{equation}
\delta\psi=\frac{3}{128\eta}v^{-5}\left[\psi_{\rm 2.5PN}v^5 + \psi_{\rm 3.5PN}v^7 + \psi_{\rm 4PN} v^8\right],
\label{eqn:delta_psi}
\end{equation}
where $v$ is the relative velocity between the inspiralling braneworld BHs, acting as the PN parameter representing the PN order at which each coefficient would appear. So we have the tidal heating contributing at three post-Newtonian orders, 2.5PN ($v^5$), 3.5PN ($v^7$) and 4PN ($v^8$) with the explicit dependence to the binary parameters as (for a derivation, see Appendix \ref{app_Phase_tidal_heating}), 
\begin{align}
&\psi_{\rm 2.5PN}=-\frac{10}{9}\,\Psi_5(3\log v + 1)~,
\\
&\psi_{\rm 3.5PN}=-\frac{5}{168}\,\Psi_5\,(952\,\eta + 995)~,
\\
&\psi_{\rm 4PN}=-\frac{20}{9}\,\left(3\, \log v - 1\right)\,\left[\Psi_5\left(F_{\rm SO}+4\pi\right)+\Psi_{8}\right]~.
\end{align}
Here $F_{\rm SO}$ is the spin-orbit coupling term (see Eq. (5) in ~\cite{Samanwaya2022}). The quantities $\Psi_{5}$ and $\Psi_{8}$ depend on the characteristic parameters associated with the BH spacetime through the relations defined below Eq.~\eqref{flux_horizon}. Notice that the binary dynamics are also influenced by each other's gravitational field, and we neglect these tidal-induced deformation effects in the phase. This is justifiable in the current scenario as the tidal deformations is a higher post-Newtonian effect (it starts appearing at the five post-Newtonian order) with a lesser contribution compared to BH absorption effect~\cite{Chakravarti:2018vlt}.

In what follows, we have incorporated these modifications to the {\tt TaylorF2} waveform model in {\tt LALSuite}~\cite{lalsuite}. {\tt TaylorF2} is an inspiral-only model for an aligned-spin binary, where the component spin angular momenta are either aligned or anti-aligned to the orbital angular momentum axis. Further, we analyse binaries of various kinds employing this waveform model after making necessary changes to the dynesty sampler implemented in {\tt Bilby}~\cite{Ashton:2018jfp}. 
We truncate the waveform model before the plunge to avoid any unmodelled effects appearing from the post-inspiral frequency region. The truncation frequency in each case is the corresponding inner-most stable circular frequency of a Kerr BH~\cite{Favata:2021vhw}, and the estimate is based on calculating the final mass and spin of the remnant BH; hence, the angular frequency considering a circular equatorial orbit around the Kerr BH, given the component masses and spins.  
\begin{figure}[t]
    \centering
    \includegraphics[width=0.5\textwidth]{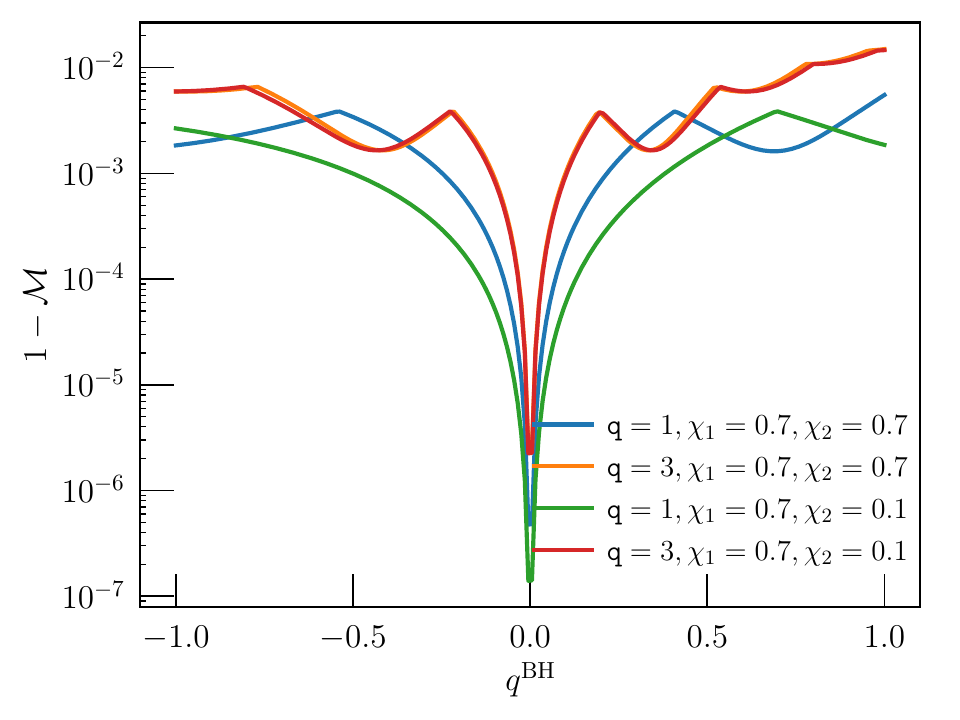}
    \caption{The mismatch between the GW waveform models corresponding to a charged BH and Kerr BH as a function of the charge parameter $q^{\rm BH}$ for different binary configurations. The binary total mass is fixed to be $20M_{\odot}$, with two different mass ratios $\mathtt{q}=(M_{1}/M_{2})=1,3$ and dimensionless spins $(\chi_{1},\chi_{2})$ taken to be $(0.7, 0.7)$ and $(0.7, 0.1)$, respectively.}
   \label{fig:mismatch}
\end{figure}

\subsection{Overview of Bayesian analysis}

In the GW data analysis, we start with the data $\rm{d}$, which contains both noise and signal. Here, noise is a random process while the signal is modelled following a particular hypothesis $\rm{H}$ and is a function of the complete set of binary parameters $\theta$. The initial prior probability distribution, $\rm{p(\theta|H)}$, restricts the range of $\theta$. If we assume that the noise is Gaussian wide-sense stationary, the likelihood function takes the form, 
\begin{equation}
    \rm{p(d|H, \theta)\propto \exp{[-\frac{(d-h|d-h)}{2}]}}~,
    \label{eq:likelihood}
\end{equation}
where $\rm{d}$ and $\rm{h}$ are the frequency domain data and signal respectively. Once we estimate the likelihood function and the prior distribution on each parameter is known, Baye's theorem provides the posterior probability distribution on each parameter as follows:
\begin{equation}
    \rm{p(\theta|H, d)=\frac{p(\theta|H)\, p(d|H, \theta)}{p(d|H)}}~.
    \label{eq:posterior}
\end{equation}
In addition, the Bayesian evidence $\rm{p(d|H)}$ is a measure of how much the data supports the hypothesis, $\rm{H}$, and is obtained by marginalizing the likelihood over the full prior volume, 
\begin{equation}
    \mathcal{Z}=\rm{p(d|H)}=\int \rm{p(\theta|H)p(d|\theta, H) d\theta}~.
\end{equation}
For the usual analyses, we assume GR accurately models the signal, and we estimate the GR evidence $\rm{\mathcal{Z}_{GR}}$. On the other hand, to check the consistency of the predictions from BHs beyond GR with the data, we include the waveform model described in Sec.~\ref{subsec:waveform} introducing a free parameter, namely the charge $q^{\rm BH}$ and obtain, $\rm{\mathcal{Z}_{nGR}}$, the evidence for non-GR signature in the data. The ratio between $\rm{\mathcal{Z}_{nGR}}$ and $\rm{\mathcal{Z}_{GR}}$ provides the Bayes factor comparing the non-GR hypothesis over the GR hypothesis. That is,
\begin{equation}
    \rm{\mathcal{B}^{nGR}_{GR} = \frac{\mathcal{Z}_{nGR}}{\mathcal{Z}_{GR}}}~.
    \label{eq:BFs}
\end{equation}
If the data is consistent with GR, the 1-dimensional marginalized posterior on $\rm{q^{BH}}$, which is defined as,
\begin{equation}
    \rm{p(q^{BH}|H_{nGR}, d)=\int p (\theta, q^{BH}|H_{nGR}, d)d\theta}
\end{equation}
will peak at zero (the GR value) and the Bayes factor $\rm{\mathcal{B}^{nGR}_{GR}}$ will be less than zero. All of these computations require evaluating the noise-weighted inner product, which has first appeared in Eq.~\ref{eq:likelihood}, and is of the form,
\begin{equation}
    (a|b)=4~\rm{Re\int_{f_{\rm low}}^{f_{\rm high}}\frac{a^{\ast}(f)\times b(f)}{Sn(f)}df}~.
    \label{eq:innerproduct}
\end{equation}
As evident, the inner product depends on the detector characteristics through the noise power spectral density $\rm{Sn(f)}$ and the lower cut-off frequency $f_{\rm low}$. In our case, we fix $f_{\rm low}$ to be $\rm{10 Hz}$ and the upper cut-off frequency $f_{\rm high}$ will vary according to the masses and spins of the binary system. To perform the posterior evaluation, we will be choosing the power spectral density corresponding to the plus era sensitivity (O5) for the LIGO detector~\cite{KAGRA:2013rdx}.

To visualise the effect of the extra phase term on the binary phasing and hence the GW waveform model, we show the mismatch between the two models in Fig.~\ref{fig:mismatch}; one includes the effect of the charge parameter ${\widetilde{h}_{\rm{nGR}}(f)}$ $({q^{\rm BH}\neq 0})$, while the other depicts a waveform model representing a binary BH in GR ${\widetilde{h}_{\rm{GR}}(f)}$ $(q^{\rm BH}= 0)$, such that,
\begin{equation}
    1-\mathcal{M}= 1 - (\widetilde{h}_{\rm{nGR}}(f),\widetilde{h}_{\rm{GR}}(f))~,
\end{equation}
where the noise weighted inner product is defined in Eq.~\ref{eq:innerproduct}. As Fig.~\ref{fig:mismatch} depicts, the mismatch between the two waveform models increases as we increase the $\rm{q^{BH}}$ value for both the mass ratios, $\mathtt{q}=(M_{1}/M_{2})=1,3$, and spins $(0.7, 0.7), (0.7, 0.1)$. The binary total mass is fixed to be $20M_{\odot}$ and ${q^{\rm BH}}$ value is chosen from $ [-1, 1]$. The mismatch for binaries with large spin and mass asymmetries is larger, indicating a better distinguishability from their Kerr BH counterparts. 

\subsection{Details of simulation}

The simulations are motivated by the findings of mismatch studies, and we sub-divide these studies into different sets focusing on the masses, spins, signal-to-noise ratios (SNRs) and the ability to identify a non-GR signature if present. To show the effect of component spins, we choose $(\chi_1, \chi_2)=(0.7, 0.6), (0.7, 0.1), (0.5, 0.3)$ and $(0.2, 0.1)$, where $\chi_1$ and $\chi_2$ are the dimensionless spins of the BHs, assumed to be aligned to the orbital angular momentum axis of the binary. Further, we consider four mass ratios, $\mathtt{q}\equiv (M_1/M_2)=1, 2, 3, 4$ to examine the effect of mass ratio on the $q^{\rm BH}$ estimate. For all these cases, the total mass is fixed to $32M_{\odot}$ and the binary is fixed at a particular location to generate a signal-to-noise ratio of $120$ in the detector. Moreover, we show posteriors on ${q^{\rm BH}}$ by choosing signal-to-noises 40 and 80 along with $120$ by varying the luminosity distance to quantify the measurability at different signal strengths. Moreover, the detectability of the tidal parameter is detailed by simulating injections with different values of ${q^{\rm BH}}$ ranging from $-0.7$ to $0.7$. 

A uniform prior range between [-1, 1] is assumed for the charge parameter for the entire analysis. Prior for the component masses are also taken uniformly between $[5, 80] M_{\odot}$ whereas for component spins uniform from [0, 0.99]. We fixed the luminosity distance, sky location (right ascension and declination), polarization angle, the inclination to the source fixed to the injected value while performing the parameter estimation analysis and verified that the ${\rm q^{BH}}$ posteriors are unaffected by this choice. While creating simulated injections, the luminosity distance has been altered according to the signal-to-noise ratio requirement. However, the right ascension, declination, and polarization angle are chosen to be 0, and the inclination to the source is fixed at 0.5 rad. 

\begin{figure}[t]
    \centering
    \includegraphics[width=0.48\textwidth]{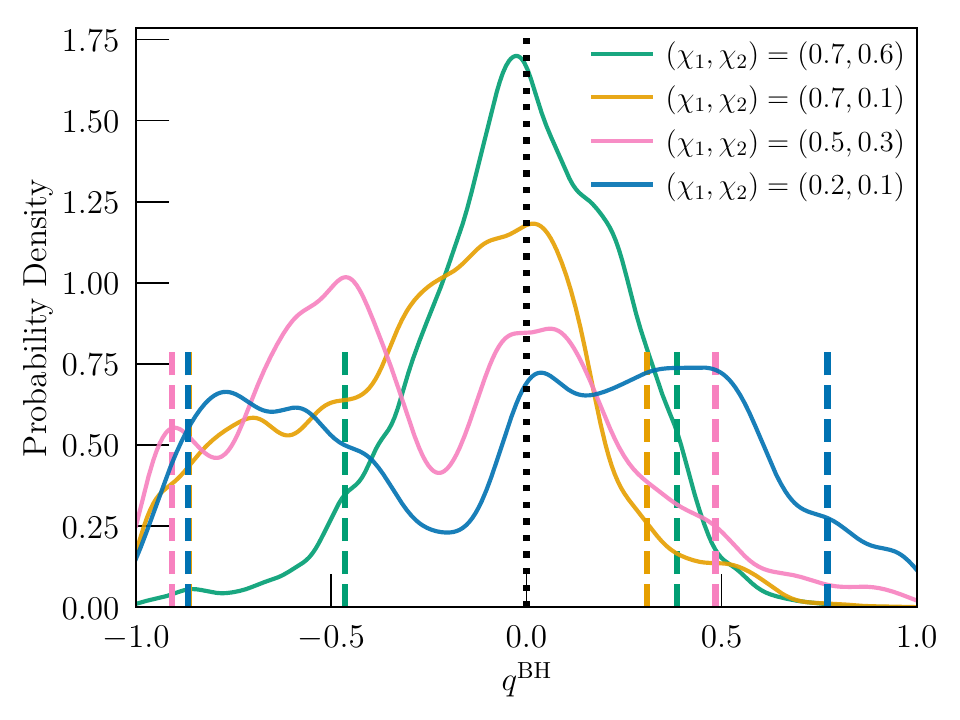}
    \caption{Posteriors on the charge parameter $q^{\rm BH}$ from simulated binary signals of total mass $32 \rm{M_{\odot}}$ and mass ratio $\mathtt{q}=(M_1/M_2)=3$ with spins $(0.7, 0.6)$, $(0.7, 0.1)$, $(0.5, 0.3)$ and $(0.2, 0.1)$. The luminosity distance to the source is chosen such that the binary generates a signal-to-noise ratio of $120$ in the detector band.}
   \label{fig:spins}
\end{figure}

It is worth mentioning that a full waveform model including inspiral-merger-ringdown regimes and spin-precession effects would be the best for our analysis. However, no such model is known in the context of braneworld and emplying the {\tt TaylorF2} waveform model is sufficient for our purpose. First of all, we have truncated the likelihood evaluation at the last stable circular frequency to ensure the validity of the PN approximation and to avoid any systematic that arises due to the presence of un-modelled physics contributing from the post-inspiral regime. Moreover, the absorption effects start to appear at 2.5 PN for spinning binaries, which is relatively lower than other effects, such as tidal deformation (a 5PN effect). Therefore, the analysis demonstrated here is largely waveform-independent and can easily be extended to more generic waveform models by simply adding the phase corrections due to tidal heating appropriately. The only non-trivial part of the above analysis corresponds to the determination of the innermost stable circular orbit frequency, for which we have used the current fits available for Kerr BHs, even for simulations with $\rm{q^{BH}}\neq0$. In the ideal case, one may use an upper cut-off frequency expression that includes the $\rm{q^{BH}}$ effect, but that is currently unavailable. Also, as emphasized above, since the test applies to inspiral-dominated signals, a slight change in the upper cut-off frequency of the analysis will most likely leave the findings unaltered. 

\section{Constraints from simulated binary signals}
\label{res1}

We generate a set of simulated binary signals (injections) in {\it zero-noise} to  study the measurability of the charge parameter $q^{\rm BH}$, including binaries of different spins, mass ratios, and signal strengths or signal-to-noise ratios. Furthermore, a set of simulations is investigated, keeping non-zero values for $q^{\rm BH}$ to quantify the detectability if present in the data. The analysis assumes that both the BHs in the binary spins are aligned/anti-aligned to the orbital angular momentum axis. 

Figure~\ref{fig:spins}, shows the posterior probability distribution on $q^{\rm BH}$ for a binary of total mass $32M_{\odot}$ and mass ratio $\mathtt{q}=(M_1/M_2)=3$. Each curve corresponds to different spin configurations, namely $(0.7, 0.6)$, $(0.7, 0.1)$, $(0.5, 0.3)$ and $(0.2, 0.1)$. Source locations of all these binaries are chosen such that the signal-to-noise ratio is $120$. The black dotted line represents the GR value ($q^{\rm BH}=0$), and the dashed lines show the 90\% bounds on $q^{\rm BH}$ for each case. It is evident from the green and orange curves of Fig. ~\ref{fig:spins} that the estimates are better when the spins are high, especially when the primary BH is largely spinning. The $q^{\rm BH}$ estimate worsens as the spin magnitudes decrease. Especially, the negative prior side is not well constrained when the spin of the secondary BH is low \footnote{For the braneworld scenario, it is expected that the tidal charge will have the same value for both the inspiralling BHs, since the origin of the charge is from higher dimension. On the other hand, in the case of electromagnetically charged BHs, unless both the BHs have some common origin, it is unlikely that they will have the same electromagnetic charge. We have done the subsequent analysis in such a case as well. However, since the number of parameters have increased, the posteriors become quite uninformative at this stage, and hence have not been described here.}. 
\begin{figure}[t]
    \centering
    \includegraphics[width=0.48\textwidth]{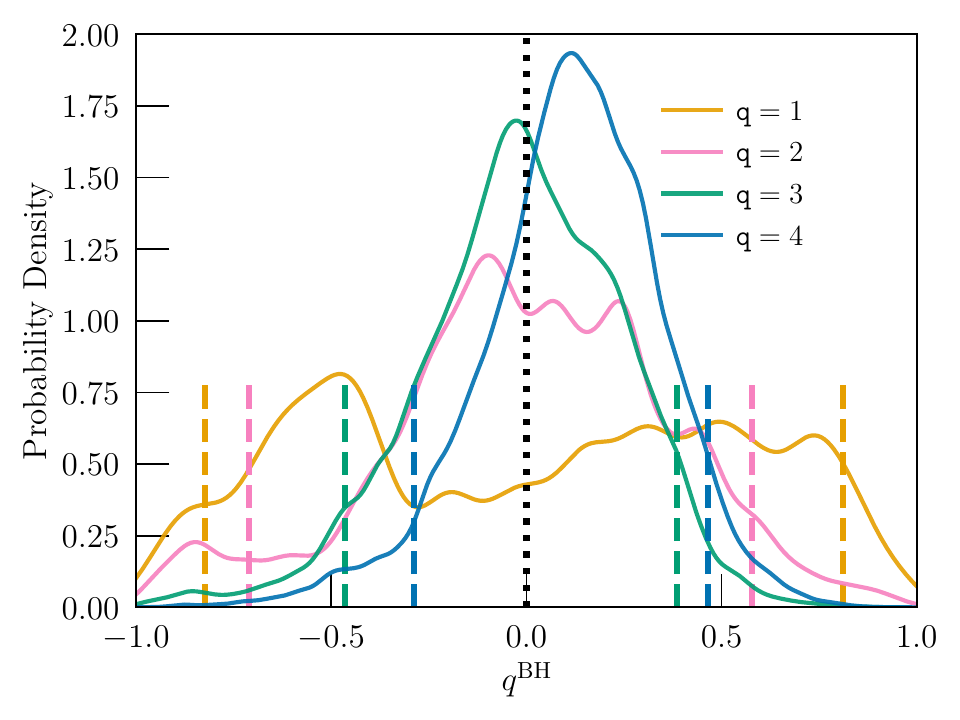}
    \caption{Posteriors on the charge parameter from simulated binary signals of total mass $32 \rm{M_{\odot}}$ and spins $(0.7, 0.6)$ with various mass ratios $\mathtt{q}=(M_1/M_2)=1, 2, 3, 4$. The signal-to-noise ratio of the signal is fixed to be $120$.}
   \label{fig:q}
\end{figure}

To study the effect of mass ratio on the estimates of the charge, we choose three mass ratios, $\mathtt{q}=(M_1/M_2)=1,2,3,4$ as shown in Fig.~\ref{fig:q}. As expected from the mismatch studies, the estimates are better as we move to larger mass ratios. Whereas equal mass binary provides the least interesting constrain on $q^{\rm BH}$. The total mass is fixed to $32M_{\odot}$ and dimensionless aligned-spin magnitudes to $(0.7, 0.6)$ and the luminosity distance and sky localization are chosen such that the signal-to-noise ratio is $120$. As evident from Fig.~\ref{fig:q}, the $90\%$ credible interval bound on the $q^{\rm BH}$ parameter is estimated to be 0.59 for mass-ratio of q=3, and the bound on $q^{\rm BH}$ is 0.4, when we consider mass-ratio to be q=4. Therefore, the bound on $q^{\rm BH}$ is stronger for the q=4 case (blue curve) than the q=3 case (green curve). This is because, the phase contributions for equal mass ratio are smaller than that of asymmetric mass ratio cases, leading to better estimates. This is consistent with earlier findings regarding BH absorption effects, See for example \cite{Maselli:2017cmm}. What is also interesting here is the slight shift in the peak of the posterior from the GR value, which is zero, with a median value of -0.13 for q=3 and 0.12 for q=4 case. This shift is arising because of the correlation between the charge parameter and the other intrinsic parameters of the binary (especially, we see that this effect is larger as the mass asymmetry of the binary increases). This implies that highly spinning binaries with large asymmetry in the mass ratio is the best candidate for detecting the existence of the charge parameter. 

\begin{figure}
    \centering
    \includegraphics[width=0.48\textwidth]{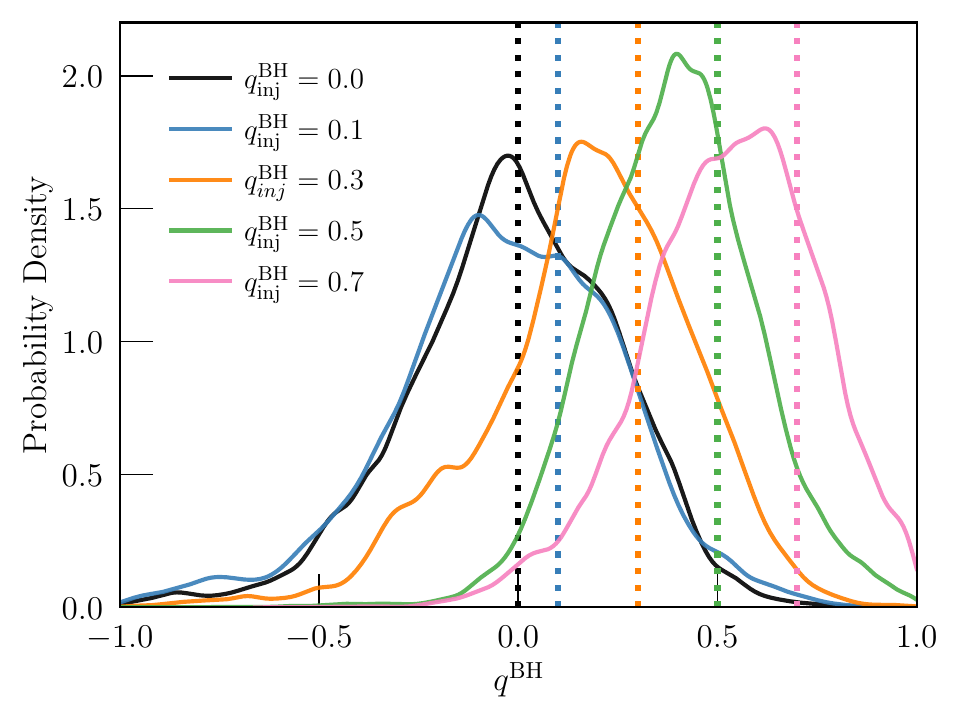}
    \includegraphics[width=0.48\textwidth]{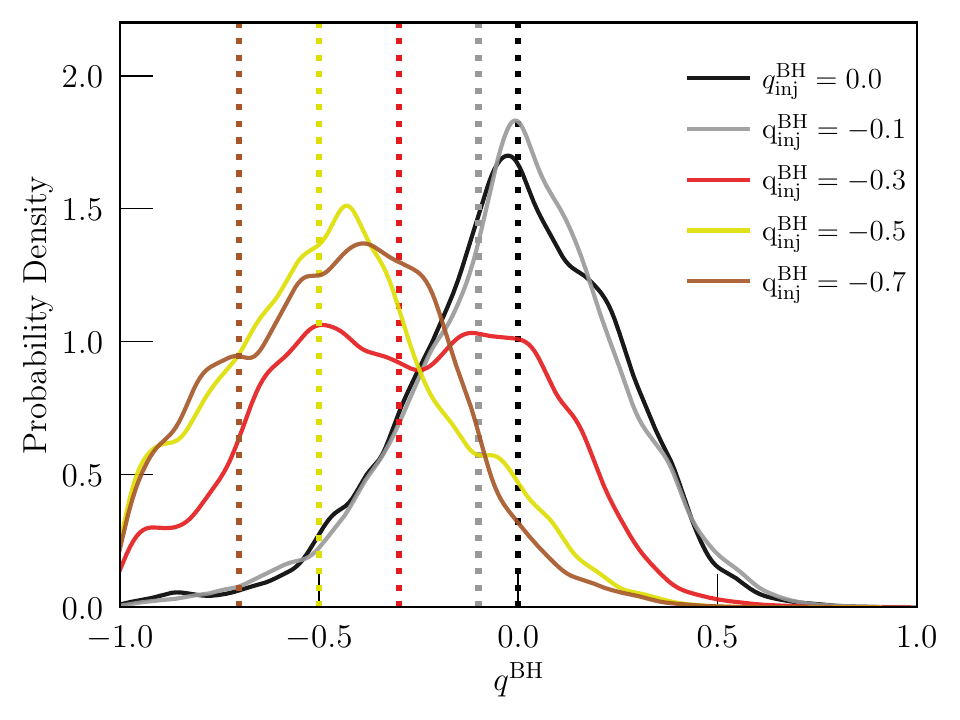}
    \caption{Posteriors on the tidal charge parameter to demonstrate the detectability of the existence of the charge parameter $q^{\rm BH}$ from the data. We choose a simulated signal from a binary BH of total mass $32 \rm{M_{\odot}}$ and mass ratio $\mathtt{q}=(M_1/M_2)=3$ with spins $(0.7, 0.6)$. The positive values of the injected charge parameter is presented in the top figure, while the negative values of the charge parameter have been presented at the bottom. The injections are marked with dotted lines and assumes a signal-to-noise ratio of $120$.}
   \label{fig:nonGR}
\end{figure}

Finally, we have performed a detailed analyses on binary BH simulations by injecting different $q^{\rm BH}$ values, namely $\rm{q^{BH}}=\pm 0.7, \pm0.5, \pm0.3, \pm0.1$. For demonstration, we fix the binary mass to be $\rm{32M_{\odot}}$, the mass ratio to be $3$, and source location and orientation in such a way that the signal produces a signal-to-noise ratio of $120$ in the advanced LIGO detector. Figure~\ref{fig:nonGR} shows the probability distribution on the charge parameter $q^{\rm BH}$ for various cases. The vertical dotted lines in Fig.~\ref{fig:nonGR} denote the initial injected values of $q^{\rm BH}$, while the central black line is the $q^{\rm BH}$ estimate, assuming the true value to be zero, i.e., the GR value. From Fig.~\ref{fig:nonGR}, it is clear that the $q^{\rm BH}$ probability distribution function shows distinct features and can be distinguished from its GR value for $|q^{\rm BH}|>0.3$. For smaller values of the charge, e.g., for $|q^{\rm BH}|\sim 0.1$, the probability distributions are indistinguishable between positive and negative values of $q^{\rm BH}$ and also with the GR value. Fig.~\ref{fig:nonGR} also indicates that the probability distribution functions are well separated for all the positive injections of $q^{\rm BH}\gtrsim 0.3$ and hence are distinguishable. While for the negative injections, though the distribution functions are different from GR value, they are indistinguishable among themselves. In other words, it is impossible to distinguish the probability distribution function for $q^{\rm BH}=-0.5$ and $q^{\rm BH}=-0.7$. Thus positive values of the charge parameter, namely those associated with extra dimensions are easier to distinguish, compared to the negative injections, associated with the electromagnetic origin. We would like to point out that there is an asymmetry between the probability distribution functions with positive and negative values of the injected tidal charge parameter. This happens because the phase change due to tidal heating depends on $q^{\rm BH}$ as well as on $(q^{\rm BH})^{2}$ (see Eqs.~(\ref{25pN}) and (\ref{5pN}) for explicit expressions), and hence $q^{\rm BH}\to -q^{\rm BH}$ is not a symmetry.

\begin{figure}[t]
    \centering
    \includegraphics[width=0.48\textwidth]{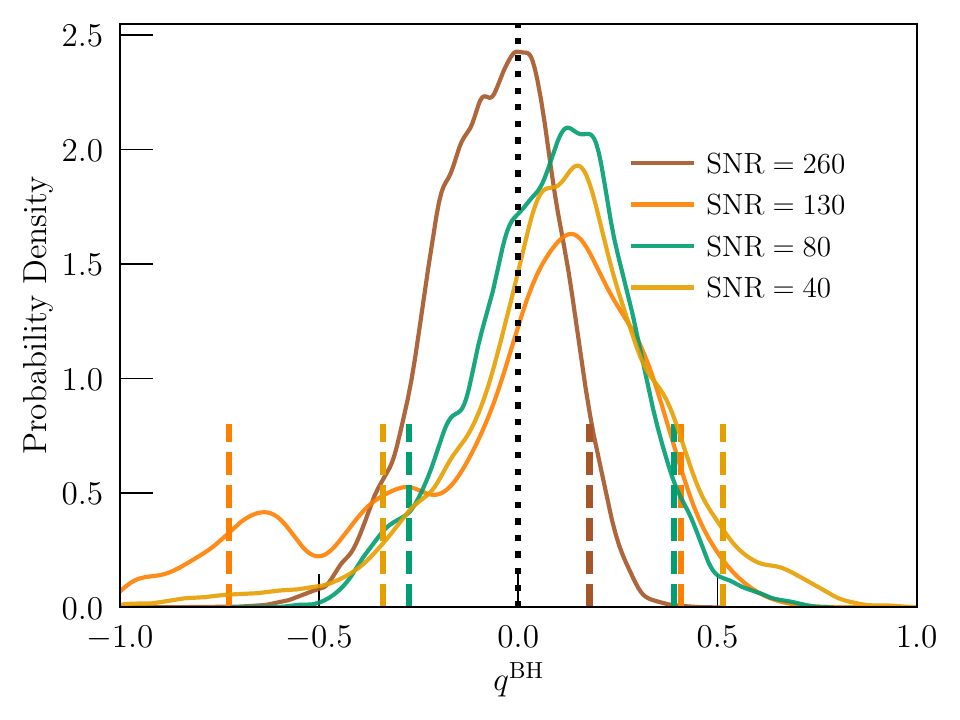}
    \caption{Posteriors on the charge parameter have been presented for different signal-to-noise ratios, to demonstrate the effect of the same on the detectability of the charge parameter. We consider a binary system with masses $(24, 8)M_{\odot}$ and spins $(0.7, 0.6)$. Four different signal-to-noise ratio values are considered, namely, $\rm{SNR}=260, 130, 80, 40$.}
   \label{fig:snr}
\end{figure}

For completeness, We have also studied the effect of the signal-to-noise ratio on the estimation of the charge parameter, and the resulting posteriors have been plotted in Fig.~\ref{fig:snr}. To demonstrate the improvement in measuring the charge parameter with respect to the signal strength, we have considered four different signal-to-noise ratios, 260, 130, 80 and 40, adjusting the distance to the source accordingly. Following our earlier conclusion, we have assumed a binary BH with asymmetric mass ratio ($\mathtt{q}=3$) and high spins ($\chi_{1}=0.7$ and $\chi_{2}=0.6$), so that the charge parameter can be better estimated. As Fig.~\ref{fig:snr} demonstrates, unless we have a high signal-to-noise ratio $\sim 200$, all possible values of $q^{\rm BH}$ are allowed, while for high signal-to-noise ratio, the distribution functions start to rule out larger values of $q^{\rm BH}$. This gives the following three criteria for observing the existence of the charge parameter --- (a) asymmetric mass ratio: more asymmetric the mass ratio, the better the chance of detection; (b) higher spins: higher is the spin of the binary BHs, better is the chance of observing them in future GW observations and (c) higher signal-to-noise ratio enhances the detection probability. With this input, we now discuss their implications for the third GW transient catalogue~\cite{ GWOSC-3} and future detectors.  

\section{Implications for the third GW transient catalog and future detectors}
\label{res2}

\begin{figure*}[]
    \centering
    \includegraphics[width=1\textwidth]{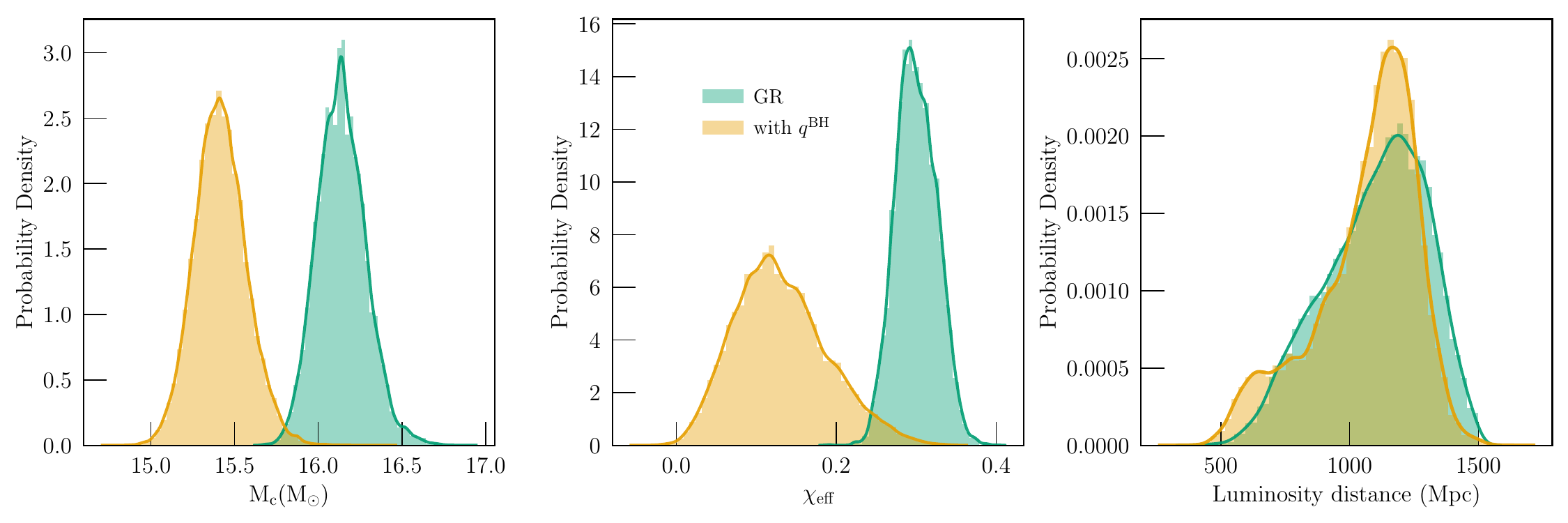}
    \caption{Posteriors on the chirp mass (a best measured component mass combination), effective spin parameter (a best measured aligned spin spin component) and luminosity distance of the GW190412 event~\cite{GW190412} have been presented, with the GR analysis and also by including the charge parameter $q^{\rm BH}$. The upper cut-off frequency is calculated for providing the maximum probable values of mass-spin posteriors obtained from the GR analysis. As evident inclusion of charge affects the intrinsic parameters, the mass and the spin, significantly.}
   \label{fig:realevent}
\end{figure*}

The GW detection from the advanced LIGO-Virgo detectors contains binary events of various types in terms of masses, spins, location, orientation and nature of the compact object. The test mentioned above for `charged' BHs suits the inspiral-dominated binaries with non-zero spins and mass asymmetry --- for instance, GW190412~\cite{GW190412}-like sources, which is one of the most asymmetric binary systems with $\sim30M_{\odot}$ primary BH and $\sim3M_{\odot}$ companion (see Tab.II of ~\cite{GW190412} for more details). Besides, the GW190412 is an inspiral-dominated event with evidence of non-zero spins, making it ideal to test the validity of the above claims against real data. 

For this purpose, we employ the non-precessing GW waveform model {\tt{TaylorF2}} with the charp mass $M_{\rm c}$, effective spin parameter $\chi_{\rm eff}$, tidal charge $q^{\rm BH}$ and the luminosity distance as free parameters, and truncate the analysis at $\rm{210Hz}$. 
The result of such an analysis is an estimation of the $q^{\rm BH}$ parameter, which reads $0.05^{0.77}_{-0.86}$ within the 90\% credible interval with respect to the mean value, $0.05$. Even though a non-zero and positive mean value is a tantalizing indication of the existence of an extra spatial dimension, the errors are very large. In particular, the GR value is well within the 90\% credible interval, and so are plenty of positive and negative values, reducing the robustness of the claim. We have also estimated the Bayes factor $\rm{\log \mathcal{B}^{nGR}_{GR}}$ supporting the non-GR hypothesis over the GR hypothesis as $\rm{\log_{e} \mathcal{B}^{nGR}_{GR}}=-6.43$. Therefore, the results look promising, while the error bar needs to be further reduced. 

Additionally, in the GW190412 analysis, we further notice that the inclusion of $q^{\rm BH}$ introduces significant shifts in the intrinsic parameters of the binary, such as masses and spins\footnote{We reiterate once again that we have not assumed the mass and spin parameters to be those obtained by fitting the GR template to the GW data. Rather, we have kept both of them, along with the luminosity distance, as free parameters and estimated along with the tidal charge $q^{\rm BH}$.}. In Fig.~\ref{fig:realevent}, we show the posteriors on the chirp mass (a combination of binary masses which is well estimated from the inspiral signal), the effective spin parameter (found to be the best representative of the aligned-spin effects of the binary) and the luminosity distance to the source. As Fig.~\ref{fig:realevent} demonstrates, the intrinsic binary parameters are significantly affected by the presence of $q^{\rm BH}$, while the extrinsic parameter remains unaltered. In particular, the percentage change in the chirp mass and the effective spin parameter becomes $(\Delta M_{\rm c}/M_{c})\sim 4.9\%$ and $(\delta \rm{\chi_{eff}}/\rm{\chi_{eff}})\sim 66.7\%$. Therefore, our results indicate that parameters beyond GR theories are highly entangled with other intrinsic parameters of the problem, particularly the spin. This makes the detection of any non-GR effect significantly challenging.  

\begin{figure*}[]
    \centering
    \includegraphics[width=0.45\textwidth]{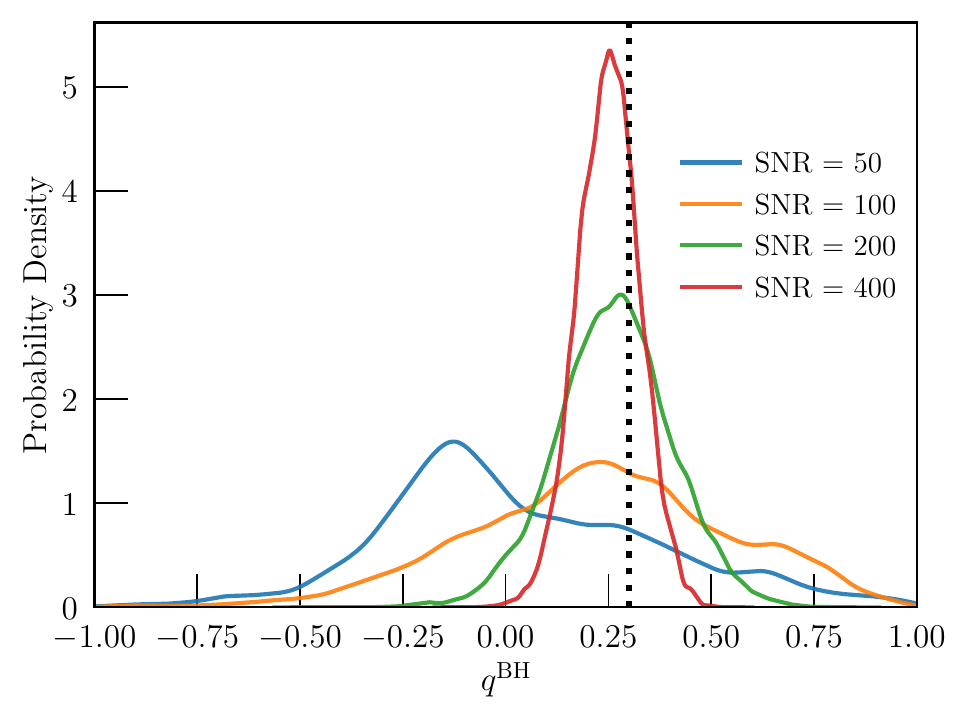}
\includegraphics[width=0.45\textwidth]{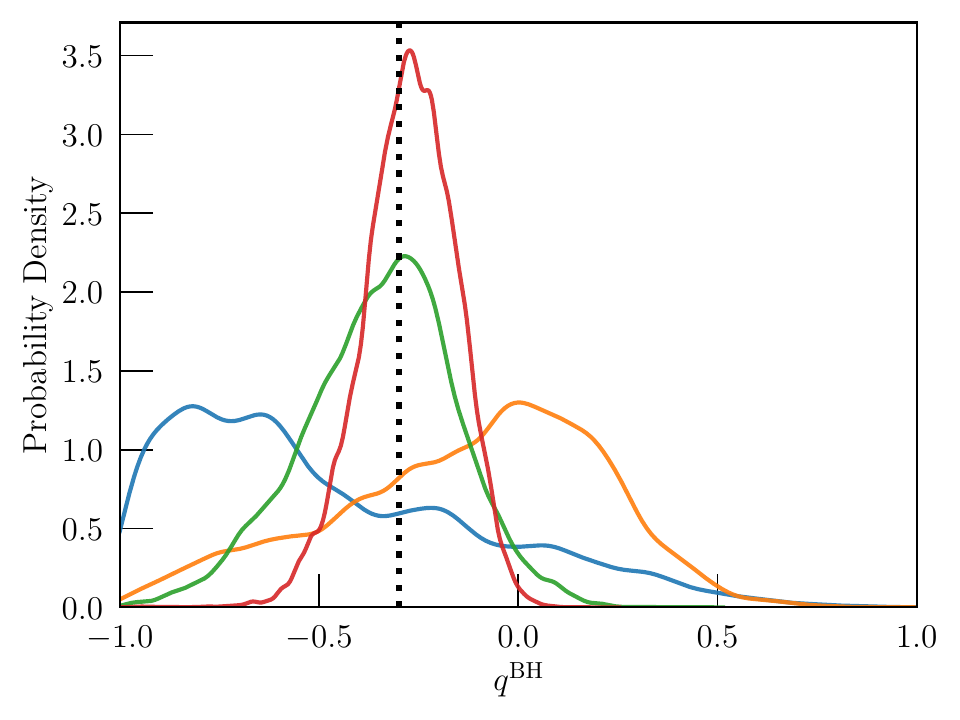}
    \caption{The probability distribution functions of the charge parameter $q_{\rm BH}$ have been presented for a binary of fixed total mass $32M_{\odot}$, mass ratio $\mathtt{q}=3$ and spins $(0.7, 0.6)$ considering four different signal-to-noise ratios $400, 200, 100$, and $50$. The plot on the left shows the distribution function for the charge with a positive injected value of $q^{\rm BH}$ ($q^{\rm BH}=0.3$), while the one on the right shows the distribution function for negative injected values of $q^{\rm BH}$ ($q^{\rm BH}=-0.3$) as indicated by the black dotted lines.}
   \label{fig:future}
\end{figure*}

We finally discuss the constraints on the charge $q^{\rm BH}$ from future GW detectors, such as Einstein Telescope and Cosmic Explorer. For this purpose we have performed the parameter analysis on simulated binaries with $200$ and $400$ as the signal-to-noise ratios, respectively and compare the constraints on $\rm{q^{BH}}$ with lower signal-to-noise ratio simulations. The corresponding posteriors on $q^{\rm BH}$, considering a simulated binary BH signal with total mass $32M_{\odot}$, and spins $(0.7, 0.6)$ are shown in Fig.~\ref{fig:future} for two possible injected values of the charge parameter, $q^{\rm BH}_{\rm inj}=\pm0.3$. As evident from Fig.~\ref{fig:future}, a signal-to-noise ratio of 200 and 400 significantly improves the detectability of the charge parameter. In particular, for signal-to-noise ratio of 400 and positive injected value of $q^{\rm BH}$ (i.e., for $q^{\rm BH}=0.3$), GR can be ruled out with more than $90\%$ confidence. While for negative $q^{\rm BH}$, the GR remains within the $90\%$ confidence interval. In summary, high signal-to-noise ratio is essential for detecting the charge parameter and ruling out GR with confidence, which requires next generation of GW detectors. 
\section{Summary}
\label{sec:summary}

We have proposed the tidal heating phenomenon in the inspiral regime of a binary BH system to be a benchmark in distinguishing binaries composed of BHs with charge from Kerr BHs using their GW signatures. For this purpose, we have started with the GW waveform model for a binary BH, parameterized in terms of an extra parameter, namely the charge, $q^{\rm BH}$. The origin for this charge can be from extra dimensions, in which case the charge is positive, or from scenarios involving simple electromagnetic interaction, in which case $q^{\rm BH}$ takes negative values. To demonstrate the measurability and detectability of the existence of such a hair from the GW observations, we have considered simulated binary BH signals of varying masses, spins and signal strength. The findings of this simulation study suggest that if the BHs are highly spinning and mass asymmetric, we will be able to perform such a distinguishability test with the advanced LIGO sensitivity since the mismatch significantly increases. Moreover, if the charge parameter is present and its value is significant, e.g., greater than $\pm 0.3$, we can detect its presence and distinguish it from GR with advanced sensitivity. Even though the current detectors are not yet sensitive enough for this test, we show that a GW190412-like event would be an ideal candidate for testing the existence of `charged' hair. Interestingly, the posteriors from the GW190412 event report a positive median value of the charge, consistent with the existence of an extra dimension, however, the error bars are huge, rendering any conclusive statement. Moreover, the Bayes factor supporting the GR hypothesis, on the other hand, was found to be $\rm{\log_{e}6.43}$, which is not a large number. This suggests that tidal heating provides an avenue to test theories beyond GR.

Besides providing simulated GW waveforms with the charge parameter and testing their distinguishability from pure GR waveforms, we have also provided a forecast analysis for the observability of the charge parameter in future GW observations. We have shown that increasing the signal-to-noise ratio considerably improves the estimation of the charge parameter and enhances its detectability. In particular, the future detector sensitivity, with a signal-to-noise ratio of 400, can significantly constrain the charge parameter. Moreover, as we have demonstrated that the tests involving charged hair of BHs will better suit more asymmetric binaries such as those in intermediate and extreme mass ratio inspirals, the future space-based GW detectors, namely DECIGO and LISA can also play a vital role in looking for the existence of non-trivial charged hair. The adaptation of the analysis presented here, in the context of the GW detectors DECIGO and LISA, will be explored in the future. 

\section*{Acknowledgments}  

S.C. and N.V.K. acknowledges discussions with Sayak Datta on several aspects of this manuscript. We also thank Khun Sang Phukon and Gregorio Carullo for reading the manuscript and providing useful comments. The research of S.C. is funded by the INSPIRE Faculty fellowship from DST, Government of India (Reg. No. DST/INSPIRE/04/2018/000893) and MATRICS grant from Science and Engineering Research Board (SERB), Government of India (Reg. No. MTR/2023/000049). N.V.K. acknowledges support from SERB for the National postdoctoral fellowship  (Reg. No. PDF/2022/000379). N.V.K. acknowledges Max Planck Computing and Data Facility computing cluster Cobra and Raven for computations. We also thank the CIT cluster provided by the LIGO Laboratory. We acknowledge National Science Foundation Grants PHY-0757058 and PHY-0823459. This material is based upon work supported by NSF's LIGO Laboratory which is a major facility fully funded by the National Science Foundation. We used the following software packages: {\tt LALSuite}~\cite{lalsuite}, {\tt bilby}~\cite{Ashton:2018jfp}, {\tt bilby\textunderscore pipe}~\cite{Romero-Shaw:2020owr}, {\tt NumPy}~\cite{2020Natur.585..357H}, {\tt PESummary}~\cite{Hoy:2020vys}, {\tt Matplotlib}~\cite{2007CSE.....9...90H}, {\tt jupyter}~\cite{soton403913}, {\tt dynesty}~\cite{speagle2020dynesty}. This document has LIGO preprint number {\tt LIGO-P2400054}.
\appendix
\begin{widetext}
\section{GW phase due to tidal heating}\label{app_Phase_tidal_heating}

The phase of the GW, as a function of velocity can be given by \cite{Tichy:1999pv}, 
\begin{align}\label{phase_integral}
\psi(v)=-2\int^{v}d\mathtt{v}\left(v^{3}-\mathtt{v}^{3}\right)\frac{(dE_{\rm orb}/d\mathtt{v})}{F_{\infty}(\mathtt{v})+F_{\rm H}(\mathtt{v})}~,
\end{align}
where, $E_{\rm orb}$ is the orbital energy of the binary BH system, $F_{\infty}$ is the GW flux at infinity and $F_{\rm H}$ is the GW flux through the BH horizon. Since we are interested in the contribution from the tidal heating alone, the following PN expansion of the orbital energy suffices for our purpose (see Eq. (3.8) of~\cite{DIS1997}) 
\begin{align}
E_{\rm orb}(\mathtt{v})=-\frac{\eta}{2}\mathtt{v}^{2}\left[1-\frac{9+\eta}{12}\mathtt{v}^{2}\right]~.
\end{align}
Such that, 
\begin{align}
\dfrac{dE_{\rm orb}}{d\mathtt{v}}=-\eta \mathtt{v}\left[1-\frac{9+\eta}{6}\mathtt{v}^{2}\right]~.
\end{align}
Similarly, for the GW flux through infinity, the following PN expansion suffices for our purpose \cite{Isoyama:2017tbp}, 
\begin{align}
F_{\infty}(\mathtt{v})&=\frac{32}{5}\eta^{2}\mathtt{v}^{10}\Big[1-\left(\frac{1247}{336}+\frac{35}{12}\eta\right)\mathtt{v}^{2}+(4\pi+F_{\rm SO})\mathtt{v}^{3}\Big]~,
\end{align}
while the GW flux through the horizon yields, 
\begin{align}
F_{\rm H}(\mathtt{v})&=\frac{32}{5}\eta^{2}\mathtt{v}^{10}\left[-\frac{\Psi_{5}}{4}\mathtt{v}^{5}+\frac{\Psi_{8}}{2}\mathtt{v}^{8}\right]~.
\end{align}
Therefore, the contribution of tidal heating from the $(1/\textrm{Flux})$ term in the phase integral, presented in Eq. (\ref{phase_integral}), yields, 
\begin{align}
\frac{1}{F_{\infty}+F_{\rm H}}&=\frac{5}{32\eta^{2}}\frac{1}{\mathtt{v}^{10}}\Bigg[\frac{\Psi_{5}}{4}\mathtt{v}^{5}
+\left(\frac{1247}{336}+\frac{35}{12}\eta\right)\frac{\Psi_{5}}{2}\mathtt{v}^{7}-\left\{\frac{\Psi_{8}}{2}+\frac{\Psi_{5}}{2}\left(4\pi+F_{\rm SO}\right)\right\}\mathtt{v}^{8}\Bigg]~.
\end{align}
Therefore the contribution of tidal heating to the phase of the GW becomes, 
\begin{align}\label{phase_integral_heating}
\psi_{\rm TH}(v)&=\frac{10}{32\eta}\int^{v}d\mathtt{v}\frac{\left(v^{3}-\mathtt{v}^{3}\right)}{\mathtt{v}^{9}}\left[1-\frac{9+\eta}{6}\mathtt{v}^{2}\right]
\Bigg[\frac{\Psi_{5}}{4}\mathtt{v}^{5}+\left(\frac{1247}{336}+\frac{35}{12}\eta\right)\frac{\Psi_{5}}{2}\mathtt{v}^{7}-\left\{\frac{\Psi_{8}}{2}+\frac{\Psi_{5}}{2}\left(4\pi+F_{\rm SO}\right)\right\}\mathtt{v}^{8}\Bigg]
\nonumber
\\
&=\frac{10}{32\eta}\int^{v}d\mathtt{v}\frac{\left(v^{3}-\mathtt{v}^{3}\right)}{\mathtt{v}^{9}}\Bigg[\frac{\Psi_{5}}{4}\mathtt{v}^{5}+\left\{2\left(\frac{1247}{336}+\frac{35}{12}\eta\right)-\frac{9+\eta}{6}\right\}\frac{\Psi_{5}}{4}\mathtt{v}^{7}-\left\{\frac{\Psi_{8}}{2}+\frac{\Psi_{5}}{2}\left(4\pi+F_{\rm SO}\right)\right\}\mathtt{v}^{8}\Bigg]
\nonumber
\\
&=\frac{10}{32\eta}\Bigg[\frac{\Psi_{5}}{4}\int^{v}d\mathtt{v}\frac{\left(v^{3}-\mathtt{v}^{3}\right)}{\mathtt{v}^{4}}
+\frac{\Psi_{5}}{4}\left(\frac{995}{168}+\frac{952}{168}\eta\right)\int^{v}d\mathtt{v}\frac{\left(v^{3}-\mathtt{v}^{3}\right)}{\mathtt{v}^{2}}
-\left\{\frac{\Psi_{8}}{2}+\frac{\Psi_{5}}{2}\left(4\pi+F_{\rm SO}\right)\right\}\int^{v}d\mathtt{v}\frac{\left(v^{3}-\mathtt{v}^{3}\right)}{\mathtt{v}}\Bigg]
\nonumber
\\
&=\frac{10}{32\eta}\Bigg[-\frac{\Psi_{5}}{12}\left(1+3\ln v\right)-\frac{3\Psi_{5}}{8}\left(\frac{995}{168}+\frac{952}{168}\eta\right)v^{2}-\left\{\frac{\Psi_{8}}{2}+\frac{\Psi_{5}}{2}\left(4\pi+F_{\rm SO}\right)\right\}\frac{v^{3}}{3}\left(3\ln v-1\right)\Bigg]~.
\end{align}
This expression has been used in the main text for computing $\delta \psi$ and identifying the 2.5 PN, 3.5 PN and 4 PN terms in the phase factor due to tidal heating. 
\end{widetext}
\bibliographystyle{apsrev}
\bibliography{ref}

\begin{thebibliography}{91}
\expandafter\ifx\csname natexlab\endcsname\relax\def\natexlab#1{#1}\fi
\expandafter\ifx\csname bibnamefont\endcsname\relax
  \def\bibnamefont#1{#1}\fi
\expandafter\ifx\csname bibfnamefont\endcsname\relax
  \def\bibfnamefont#1{#1}\fi
\expandafter\ifx\csname citenamefont\endcsname\relax
  \def\citenamefont#1{#1}\fi
\expandafter\ifx\csname url\endcsname\relax
  \def\url#1{\texttt{#1}}\fi
\expandafter\ifx\csname urlprefix\endcsname\relax\def\urlprefix{URL }\fi
\providecommand{\bibinfo}[2]{#2}
\providecommand{\eprint}[2][]{\url{#2}}

\bibitem[{\citenamefont{Aasi et~al.}(2015)}]{AdvLIGO}
\bibinfo{author}{\bibfnamefont{J.}~\bibnamefont{Aasi}} \bibnamefont{et~al.}
  (\bibinfo{collaboration}{LIGO Scientific}), \bibinfo{journal}{Class. Quant.
  Grav.} \textbf{\bibinfo{volume}{32}}, \bibinfo{pages}{074001}
  (\bibinfo{year}{2015}), \eprint{1411.4547}.

\bibitem[{\citenamefont{Acernese et~al.}(2015)}]{AdvVirgo}
\bibinfo{author}{\bibfnamefont{F.}~\bibnamefont{Acernese}} \bibnamefont{et~al.}
  (\bibinfo{collaboration}{VIRGO}), \bibinfo{journal}{Class. Quant. Grav.}
  \textbf{\bibinfo{volume}{32}}, \bibinfo{pages}{024001}
  (\bibinfo{year}{2015}), \eprint{1408.3978}.

\bibitem[{vir()}]{virgo}
\bibinfo{howpublished}{\url{http://www.virgo.infn.it}}.

\bibitem[{lig()}]{ligo}
\bibinfo{howpublished}{\url{http://www.ligo.caltech.edu}}.

\bibitem[{\citenamefont{Abbott et~al.}(2021{\natexlab{a}})}]{GWTC3-catalog}
\bibinfo{author}{\bibfnamefont{R.}~\bibnamefont{Abbott}} \bibnamefont{et~al.}
  (\bibinfo{collaboration}{LIGO Scientific, VIRGO, KAGRA})
  (\bibinfo{year}{2021}{\natexlab{a}}), \eprint{2111.03606}.

\bibitem[{\citenamefont{Abbott et~al.}(2021{\natexlab{b}})}]{GWTC-2-catalog}
\bibinfo{author}{\bibfnamefont{R.}~\bibnamefont{Abbott}} \bibnamefont{et~al.}
  (\bibinfo{collaboration}{LIGO Scientific, Virgo}), \bibinfo{journal}{Phys.
  Rev. X} \textbf{\bibinfo{volume}{11}}, \bibinfo{pages}{021053}
  (\bibinfo{year}{2021}{\natexlab{b}}), \eprint{2010.14527}.

\bibitem[{\citenamefont{Abbott et~al.}(2021{\natexlab{c}})}]{GWTC-2.1-catalog}
\bibinfo{author}{\bibfnamefont{R.}~\bibnamefont{Abbott}} \bibnamefont{et~al.}
  (\bibinfo{collaboration}{LIGO Scientific, VIRGO})
  (\bibinfo{year}{2021}{\natexlab{c}}), \eprint{2108.01045}.

\bibitem[{\citenamefont{Nitz et~al.}(2023)\citenamefont{Nitz, Kumar, Wang,
  Kastha, Wu, Sch\"afer, Dhurkunde, and Capano}}]{4OGC}
\bibinfo{author}{\bibfnamefont{A.~H.} \bibnamefont{Nitz}},
  \bibinfo{author}{\bibfnamefont{S.}~\bibnamefont{Kumar}},
  \bibinfo{author}{\bibfnamefont{Y.-F.} \bibnamefont{Wang}},
  \bibinfo{author}{\bibfnamefont{S.}~\bibnamefont{Kastha}},
  \bibinfo{author}{\bibfnamefont{S.}~\bibnamefont{Wu}},
  \bibinfo{author}{\bibfnamefont{M.}~\bibnamefont{Sch\"afer}},
  \bibinfo{author}{\bibfnamefont{R.}~\bibnamefont{Dhurkunde}},
  \bibnamefont{and} \bibinfo{author}{\bibfnamefont{C.~D.}
  \bibnamefont{Capano}}, \bibinfo{journal}{Astrophys. J.}
  \textbf{\bibinfo{volume}{946}}, \bibinfo{pages}{59} (\bibinfo{year}{2023}),
  \eprint{2112.06878}.

\bibitem[{\citenamefont{Mehta et~al.}(2023)\citenamefont{Mehta, Olsen, Wadekar,
  Roulet, Venumadhav, Mushkin, Zackay, and Zaldarriaga}}]{IAS-3}
\bibinfo{author}{\bibfnamefont{A.~K.} \bibnamefont{Mehta}},
  \bibinfo{author}{\bibfnamefont{S.}~\bibnamefont{Olsen}},
  \bibinfo{author}{\bibfnamefont{D.}~\bibnamefont{Wadekar}},
  \bibinfo{author}{\bibfnamefont{J.}~\bibnamefont{Roulet}},
  \bibinfo{author}{\bibfnamefont{T.}~\bibnamefont{Venumadhav}},
  \bibinfo{author}{\bibfnamefont{J.}~\bibnamefont{Mushkin}},
  \bibinfo{author}{\bibfnamefont{B.}~\bibnamefont{Zackay}}, \bibnamefont{and}
  \bibinfo{author}{\bibfnamefont{M.}~\bibnamefont{Zaldarriaga}}
  (\bibinfo{year}{2023}), \eprint{2311.06061}.

\bibitem[{\citenamefont{Rich Abbott et~al.}(2021)\citenamefont{Rich Abbott,
  Thomas D. Abbott et~al.}}]{RICHABBOTT2021100658}
\bibinfo{author}{\bibnamefont{Rich Abbott}},
  \bibinfo{author}{\bibnamefont{Thomas D. Abbott}}, \bibnamefont{et~al.},
  \bibinfo{journal}{SoftwareX} \textbf{\bibinfo{volume}{13}},
  \bibinfo{pages}{100658} (\bibinfo{year}{2021}), ISSN
  \bibinfo{issn}{2352-7110},
  \urlprefix\url{https://www.sciencedirect.com/science/article/pii/S2352711021000030}.

\bibitem[{\citenamefont{Abbott et~al.}(2017)}]{GW170817}
\bibinfo{author}{\bibfnamefont{B.~P.} \bibnamefont{Abbott}}
  \bibnamefont{et~al.} (\bibinfo{collaboration}{LIGO Scientific, Virgo}),
  \bibinfo{journal}{Phys. Rev. Lett.} \textbf{\bibinfo{volume}{119}},
  \bibinfo{pages}{161101} (\bibinfo{year}{2017}), \eprint{1710.05832}.

\bibitem[{\citenamefont{Abbott et~al.}(2020{\natexlab{a}})}]{GW190425}
\bibinfo{author}{\bibfnamefont{B.~P.} \bibnamefont{Abbott}}
  \bibnamefont{et~al.} (\bibinfo{collaboration}{LIGO Scientific, Virgo}),
  \bibinfo{journal}{Astrophys. J. Lett.} \textbf{\bibinfo{volume}{892}},
  \bibinfo{pages}{L3} (\bibinfo{year}{2020}{\natexlab{a}}),
  \eprint{2001.01761}.

\bibitem[{\citenamefont{Abbott et~al.}(2020{\natexlab{b}})}]{GW190521}
\bibinfo{author}{\bibfnamefont{R.}~\bibnamefont{Abbott}} \bibnamefont{et~al.}
  (\bibinfo{collaboration}{LIGO Scientific, Virgo}), \bibinfo{journal}{Phys.
  Rev. Lett.} \textbf{\bibinfo{volume}{125}}, \bibinfo{pages}{101102}
  (\bibinfo{year}{2020}{\natexlab{b}}), \eprint{2009.01075}.

\bibitem[{\citenamefont{Abbott
  et~al.}(2021{\natexlab{d}})}]{NS_BH_GW200105_GW200115}
\bibinfo{author}{\bibfnamefont{R.}~\bibnamefont{Abbott}} \bibnamefont{et~al.}
  (\bibinfo{collaboration}{LIGO Scientific, KAGRA, VIRGO}),
  \bibinfo{journal}{Astrophys. J. Lett.} \textbf{\bibinfo{volume}{915}},
  \bibinfo{pages}{L5} (\bibinfo{year}{2021}{\natexlab{d}}),
  \eprint{2106.15163}.

\bibitem[{\citenamefont{Abbott et~al.}(2019)}]{GWTC-1-TGR}
\bibinfo{author}{\bibfnamefont{B.~P.} \bibnamefont{Abbott}}
  \bibnamefont{et~al.} (\bibinfo{collaboration}{LIGO Scientific, Virgo}),
  \bibinfo{journal}{Phys. Rev. D} \textbf{\bibinfo{volume}{100}},
  \bibinfo{pages}{104036} (\bibinfo{year}{2019}), \eprint{1903.04467}.

\bibitem[{\citenamefont{Abbott et~al.}(2021{\natexlab{e}})}]{GWTC-2-TGR}
\bibinfo{author}{\bibfnamefont{R.}~\bibnamefont{Abbott}} \bibnamefont{et~al.}
  (\bibinfo{collaboration}{LIGO Scientific, Virgo}), \bibinfo{journal}{Phys.
  Rev. X} \textbf{\bibinfo{volume}{11}}, \bibinfo{pages}{021053}
  (\bibinfo{year}{2021}{\natexlab{e}}), \eprint{2010.14527}.

\bibitem[{\citenamefont{Abbott et~al.}(2021{\natexlab{f}})}]{GWTC-3-TGR}
\bibinfo{author}{\bibfnamefont{R.}~\bibnamefont{Abbott}} \bibnamefont{et~al.}
  (\bibinfo{collaboration}{LIGO Scientific, VIRGO, KAGRA})
  (\bibinfo{year}{2021}{\natexlab{f}}), \eprint{2111.03606}.

\bibitem[{\citenamefont{Abbott et~al.}(2016)}]{TOGGW150914}
\bibinfo{author}{\bibfnamefont{B.~P.} \bibnamefont{Abbott}}
  \bibnamefont{et~al.} (\bibinfo{collaboration}{Virgo, LIGO Scientific}),
  \bibinfo{journal}{Phys. Rev. Lett.} \textbf{\bibinfo{volume}{116}},
  \bibinfo{pages}{221101} (\bibinfo{year}{2016}), \eprint{1602.03841}.

\bibitem[{\citenamefont{Berti et~al.}(2015)\citenamefont{Berti, Barausse,
  Cardoso, Gualtieri, Pani, Sperhake, Stein, Wex, Yagi, Baker
  et~al.}}]{Berti_2015}
\bibinfo{author}{\bibfnamefont{E.}~\bibnamefont{Berti}},
  \bibinfo{author}{\bibfnamefont{E.}~\bibnamefont{Barausse}},
  \bibinfo{author}{\bibfnamefont{V.}~\bibnamefont{Cardoso}},
  \bibinfo{author}{\bibfnamefont{L.}~\bibnamefont{Gualtieri}},
  \bibinfo{author}{\bibfnamefont{P.}~\bibnamefont{Pani}},
  \bibinfo{author}{\bibfnamefont{U.}~\bibnamefont{Sperhake}},
  \bibinfo{author}{\bibfnamefont{L.~C.} \bibnamefont{Stein}},
  \bibinfo{author}{\bibfnamefont{N.}~\bibnamefont{Wex}},
  \bibinfo{author}{\bibfnamefont{K.}~\bibnamefont{Yagi}},
  \bibinfo{author}{\bibfnamefont{T.}~\bibnamefont{Baker}},
  \bibnamefont{et~al.}, \bibinfo{journal}{Classical and Quantum Gravity}
  \textbf{\bibinfo{volume}{32}}, \bibinfo{pages}{243001}
  (\bibinfo{year}{2015}),
  \urlprefix\url{https://dx.doi.org/10.1088/0264-9381/32/24/243001}.

\bibitem[{\citenamefont{Cardoso et~al.}(2017)\citenamefont{Cardoso, Franzin,
  Maselli, Pani, and Raposo}}]{Cardoso:2017cfl}
\bibinfo{author}{\bibfnamefont{V.}~\bibnamefont{Cardoso}},
  \bibinfo{author}{\bibfnamefont{E.}~\bibnamefont{Franzin}},
  \bibinfo{author}{\bibfnamefont{A.}~\bibnamefont{Maselli}},
  \bibinfo{author}{\bibfnamefont{P.}~\bibnamefont{Pani}}, \bibnamefont{and}
  \bibinfo{author}{\bibfnamefont{G.}~\bibnamefont{Raposo}},
  \bibinfo{journal}{Phys. Rev. D} \textbf{\bibinfo{volume}{95}},
  \bibinfo{pages}{084014} (\bibinfo{year}{2017}), \bibinfo{note}{[Addendum:
  Phys.Rev.D 95, 089901 (2017)]}, \eprint{1701.01116}.

\bibitem[{\citenamefont{Psaltis et~al.}(2020)}]{EventHorizonTelescope2020TGR}
\bibinfo{author}{\bibfnamefont{D.}~\bibnamefont{Psaltis}} \bibnamefont{et~al.}
  (\bibinfo{collaboration}{Event Horizon Telescope}), \bibinfo{journal}{Phys.
  Rev. Lett.} \textbf{\bibinfo{volume}{125}}, \bibinfo{pages}{141104}
  (\bibinfo{year}{2020}), \eprint{2010.01055}.

\bibitem[{\citenamefont{Aliev and Gumrukcuoglu}(2005)}]{Aliev:2005bi}
\bibinfo{author}{\bibfnamefont{A.~N.} \bibnamefont{Aliev}} \bibnamefont{and}
  \bibinfo{author}{\bibfnamefont{A.~E.} \bibnamefont{Gumrukcuoglu}},
  \bibinfo{journal}{Phys. Rev. D} \textbf{\bibinfo{volume}{71}},
  \bibinfo{pages}{104027} (\bibinfo{year}{2005}), \eprint{hep-th/0502223}.

\bibitem[{\citenamefont{Harko and Mak}(2004)}]{Harko:2004ui}
\bibinfo{author}{\bibfnamefont{T.}~\bibnamefont{Harko}} \bibnamefont{and}
  \bibinfo{author}{\bibfnamefont{M.~K.} \bibnamefont{Mak}},
  \bibinfo{journal}{Phys. Rev. D} \textbf{\bibinfo{volume}{69}},
  \bibinfo{pages}{064020} (\bibinfo{year}{2004}), \eprint{gr-qc/0401049}.

\bibitem[{\citenamefont{Dadhich et~al.}(2000)\citenamefont{Dadhich, Maartens,
  Papadopoulos, and Rezania}}]{Dadhich:2000am}
\bibinfo{author}{\bibfnamefont{N.}~\bibnamefont{Dadhich}},
  \bibinfo{author}{\bibfnamefont{R.}~\bibnamefont{Maartens}},
  \bibinfo{author}{\bibfnamefont{P.}~\bibnamefont{Papadopoulos}},
  \bibnamefont{and} \bibinfo{author}{\bibfnamefont{V.}~\bibnamefont{Rezania}},
  \bibinfo{journal}{Phys. Lett. B} \textbf{\bibinfo{volume}{487}},
  \bibinfo{pages}{1} (\bibinfo{year}{2000}), \eprint{hep-th/0003061}.

\bibitem[{\citenamefont{Shiromizu et~al.}(2000)\citenamefont{Shiromizu, Maeda,
  and Sasaki}}]{Shiromizu:1999wj}
\bibinfo{author}{\bibfnamefont{T.}~\bibnamefont{Shiromizu}},
  \bibinfo{author}{\bibfnamefont{K.-i.} \bibnamefont{Maeda}}, \bibnamefont{and}
  \bibinfo{author}{\bibfnamefont{M.}~\bibnamefont{Sasaki}},
  \bibinfo{journal}{Phys. Rev. D} \textbf{\bibinfo{volume}{62}},
  \bibinfo{pages}{024012} (\bibinfo{year}{2000}), \eprint{gr-qc/9910076}.

\bibitem[{\citenamefont{Babichev and Charmousis}(2014)}]{Babichev:2013cya}
\bibinfo{author}{\bibfnamefont{E.}~\bibnamefont{Babichev}} \bibnamefont{and}
  \bibinfo{author}{\bibfnamefont{C.}~\bibnamefont{Charmousis}},
  \bibinfo{journal}{JHEP} \textbf{\bibinfo{volume}{08}}, \bibinfo{pages}{106}
  (\bibinfo{year}{2014}), \eprint{1312.3204}.

\bibitem[{\citenamefont{Barrientos et~al.}(2017)\citenamefont{Barrientos,
  Cordonier-Tello, Izaurieta, Medina, Narbona, Rodr\'\i{}guez, and
  Valdivia}}]{Barrientos:2017utp}
\bibinfo{author}{\bibfnamefont{J.}~\bibnamefont{Barrientos}},
  \bibinfo{author}{\bibfnamefont{F.}~\bibnamefont{Cordonier-Tello}},
  \bibinfo{author}{\bibfnamefont{F.}~\bibnamefont{Izaurieta}},
  \bibinfo{author}{\bibfnamefont{P.}~\bibnamefont{Medina}},
  \bibinfo{author}{\bibfnamefont{D.}~\bibnamefont{Narbona}},
  \bibinfo{author}{\bibfnamefont{E.}~\bibnamefont{Rodr\'\i{}guez}},
  \bibnamefont{and} \bibinfo{author}{\bibfnamefont{O.}~\bibnamefont{Valdivia}},
  \bibinfo{journal}{Phys. Rev. D} \textbf{\bibinfo{volume}{96}},
  \bibinfo{pages}{084023} (\bibinfo{year}{2017}), \eprint{1703.09686}.

\bibitem[{\citenamefont{Babichev et~al.}(2015)\citenamefont{Babichev,
  Charmousis, and Hassaine}}]{Babichev:2015rva}
\bibinfo{author}{\bibfnamefont{E.}~\bibnamefont{Babichev}},
  \bibinfo{author}{\bibfnamefont{C.}~\bibnamefont{Charmousis}},
  \bibnamefont{and} \bibinfo{author}{\bibfnamefont{M.}~\bibnamefont{Hassaine}},
  \bibinfo{journal}{JCAP} \textbf{\bibinfo{volume}{05}}, \bibinfo{pages}{031}
  (\bibinfo{year}{2015}), \eprint{1503.02545}.

\bibitem[{\citenamefont{Maeda and Dadhich}(2007)}]{Maeda:2006hj}
\bibinfo{author}{\bibfnamefont{H.}~\bibnamefont{Maeda}} \bibnamefont{and}
  \bibinfo{author}{\bibfnamefont{N.}~\bibnamefont{Dadhich}},
  \bibinfo{journal}{Phys. Rev. D} \textbf{\bibinfo{volume}{75}},
  \bibinfo{pages}{044007} (\bibinfo{year}{2007}), \eprint{hep-th/0611188}.

\bibitem[{\citenamefont{Faraoni and Capozziello}(2011)}]{Faraoni:2010pgm}
\bibinfo{author}{\bibfnamefont{V.}~\bibnamefont{Faraoni}} \bibnamefont{and}
  \bibinfo{author}{\bibfnamefont{S.}~\bibnamefont{Capozziello}},
  \emph{\bibinfo{title}{{Beyond Einstein Gravity}: {A Survey of Gravitational
  Theories for Cosmology and Astrophysics}}} (\bibinfo{publisher}{Springer},
  \bibinfo{address}{Dordrecht}, \bibinfo{year}{2011}), ISBN
  \bibinfo{isbn}{978-94-007-0164-9, 978-94-007-0165-6}.

\bibitem[{\citenamefont{Capozziello et~al.}(2013)\citenamefont{Capozziello,
  Gonzalez, Saridakis, and Vasquez}}]{Capozziello:2012zj}
\bibinfo{author}{\bibfnamefont{S.}~\bibnamefont{Capozziello}},
  \bibinfo{author}{\bibfnamefont{P.~A.} \bibnamefont{Gonzalez}},
  \bibinfo{author}{\bibfnamefont{E.~N.} \bibnamefont{Saridakis}},
  \bibnamefont{and} \bibinfo{author}{\bibfnamefont{Y.}~\bibnamefont{Vasquez}},
  \bibinfo{journal}{JHEP} \textbf{\bibinfo{volume}{02}}, \bibinfo{pages}{039}
  (\bibinfo{year}{2013}), \eprint{1210.1098}.

\bibitem[{\citenamefont{Hughes}(2000)}]{Hughes:1999bq}
\bibinfo{author}{\bibfnamefont{S.~A.} \bibnamefont{Hughes}},
  \bibinfo{journal}{Phys. Rev. D} \textbf{\bibinfo{volume}{61}},
  \bibinfo{pages}{084004} (\bibinfo{year}{2000}), \bibinfo{note}{[Erratum:
  Phys.Rev.D 63, 049902 (2001), Erratum: Phys.Rev.D 65, 069902 (2002), Erratum:
  Phys.Rev.D 67, 089901 (2003), Erratum: Phys.Rev.D 78, 109902 (2008), Erratum:
  Phys.Rev.D 90, 109904 (2014)]}, \eprint{gr-qc/9910091}.

\bibitem[{\citenamefont{Hartle}(1973)}]{Hartle1973}
\bibinfo{author}{\bibfnamefont{J.~B.} \bibnamefont{Hartle}},
  \bibinfo{journal}{Phys. Rev. D} \textbf{\bibinfo{volume}{8}},
  \bibinfo{pages}{1010} (\bibinfo{year}{1973}),
  \urlprefix\url{https://link.aps.org/doi/10.1103/PhysRevD.8.1010}.

\bibitem[{\citenamefont{Isoyama and Nakano}(2018)}]{Isoyama:2017tbp}
\bibinfo{author}{\bibfnamefont{S.}~\bibnamefont{Isoyama}} \bibnamefont{and}
  \bibinfo{author}{\bibfnamefont{H.}~\bibnamefont{Nakano}},
  \bibinfo{journal}{Class. Quant. Grav.} \textbf{\bibinfo{volume}{35}},
  \bibinfo{pages}{024001} (\bibinfo{year}{2018}), \eprint{1705.03869}.

\bibitem[{\citenamefont{Chatziioannou et~al.}(2013)\citenamefont{Chatziioannou,
  Poisson, and Yunes}}]{Chatziioannou:2012gq}
\bibinfo{author}{\bibfnamefont{K.}~\bibnamefont{Chatziioannou}},
  \bibinfo{author}{\bibfnamefont{E.}~\bibnamefont{Poisson}}, \bibnamefont{and}
  \bibinfo{author}{\bibfnamefont{N.}~\bibnamefont{Yunes}},
  \bibinfo{journal}{Phys. Rev. D} \textbf{\bibinfo{volume}{87}},
  \bibinfo{pages}{044022} (\bibinfo{year}{2013}), \eprint{1211.1686}.

\bibitem[{\citenamefont{Datta et~al.}(2020)\citenamefont{Datta, Brito, Bose,
  Pani, and Hughes}}]{Datta:2019epe}
\bibinfo{author}{\bibfnamefont{S.}~\bibnamefont{Datta}},
  \bibinfo{author}{\bibfnamefont{R.}~\bibnamefont{Brito}},
  \bibinfo{author}{\bibfnamefont{S.}~\bibnamefont{Bose}},
  \bibinfo{author}{\bibfnamefont{P.}~\bibnamefont{Pani}}, \bibnamefont{and}
  \bibinfo{author}{\bibfnamefont{S.~A.} \bibnamefont{Hughes}},
  \bibinfo{journal}{Phys. Rev. D} \textbf{\bibinfo{volume}{101}},
  \bibinfo{pages}{044004} (\bibinfo{year}{2020}), \eprint{1910.07841}.

\bibitem[{\citenamefont{Datta}(2020)}]{Datta:2020rvo}
\bibinfo{author}{\bibfnamefont{S.}~\bibnamefont{Datta}},
  \bibinfo{journal}{Phys. Rev. D} \textbf{\bibinfo{volume}{102}},
  \bibinfo{pages}{064040} (\bibinfo{year}{2020}), \eprint{2002.04480}.

\bibitem[{\citenamefont{Datta and Bose}(2019)}]{Datta:2019euh}
\bibinfo{author}{\bibfnamefont{S.}~\bibnamefont{Datta}} \bibnamefont{and}
  \bibinfo{author}{\bibfnamefont{S.}~\bibnamefont{Bose}},
  \bibinfo{journal}{Phys. Rev. D} \textbf{\bibinfo{volume}{99}},
  \bibinfo{pages}{084001} (\bibinfo{year}{2019}), \eprint{1902.01723}.

\bibitem[{\citenamefont{Alvi}(2001)}]{Alvi:2001mx}
\bibinfo{author}{\bibfnamefont{K.}~\bibnamefont{Alvi}}, \bibinfo{journal}{Phys.
  Rev. D} \textbf{\bibinfo{volume}{64}}, \bibinfo{pages}{104020}
  (\bibinfo{year}{2001}), \eprint{gr-qc/0107080}.

\bibitem[{\citenamefont{Poisson}(2004)}]{Poisson:2004cw}
\bibinfo{author}{\bibfnamefont{E.}~\bibnamefont{Poisson}},
  \bibinfo{journal}{Phys. Rev. D} \textbf{\bibinfo{volume}{70}},
  \bibinfo{pages}{084044} (\bibinfo{year}{2004}), \eprint{gr-qc/0407050}.

\bibitem[{\citenamefont{Saketh et~al.}(2023)\citenamefont{Saketh, Steinhoff,
  Vines, and Buonanno}}]{Saketh:2022xjb}
\bibinfo{author}{\bibfnamefont{M.~V.~S.} \bibnamefont{Saketh}},
  \bibinfo{author}{\bibfnamefont{J.}~\bibnamefont{Steinhoff}},
  \bibinfo{author}{\bibfnamefont{J.}~\bibnamefont{Vines}}, \bibnamefont{and}
  \bibinfo{author}{\bibfnamefont{A.}~\bibnamefont{Buonanno}},
  \bibinfo{journal}{Phys. Rev. D} \textbf{\bibinfo{volume}{107}},
  \bibinfo{pages}{084006} (\bibinfo{year}{2023}), \eprint{2212.13095}.

\bibitem[{\citenamefont{Datta}(2023)}]{Datta:2023wsn}
\bibinfo{author}{\bibfnamefont{S.}~\bibnamefont{Datta}} (\bibinfo{year}{2023}),
  \eprint{2305.03771}.

\bibitem[{\citenamefont{Mukherjee et~al.}(2023)\citenamefont{Mukherjee, Datta,
  Phukon, and Bose}}]{Mukherjee:2023pge}
\bibinfo{author}{\bibfnamefont{S.}~\bibnamefont{Mukherjee}},
  \bibinfo{author}{\bibfnamefont{S.}~\bibnamefont{Datta}},
  \bibinfo{author}{\bibfnamefont{K.~S.} \bibnamefont{Phukon}},
  \bibnamefont{and} \bibinfo{author}{\bibfnamefont{S.}~\bibnamefont{Bose}}
  (\bibinfo{year}{2023}), \eprint{2311.17554}.

\bibitem[{\citenamefont{Datta et~al.}(2021)\citenamefont{Datta, Phukon, and
  Bose}}]{Datta:2020gem}
\bibinfo{author}{\bibfnamefont{S.}~\bibnamefont{Datta}},
  \bibinfo{author}{\bibfnamefont{K.~S.} \bibnamefont{Phukon}},
  \bibnamefont{and} \bibinfo{author}{\bibfnamefont{S.}~\bibnamefont{Bose}},
  \bibinfo{journal}{Phys. Rev. D} \textbf{\bibinfo{volume}{104}},
  \bibinfo{pages}{084006} (\bibinfo{year}{2021}), \eprint{2004.05974}.

\bibitem[{\citenamefont{Maselli et~al.}(2018)\citenamefont{Maselli, Pani,
  Cardoso, Abdelsalhin, Gualtieri, and Ferrari}}]{Maselli:2017cmm}
\bibinfo{author}{\bibfnamefont{A.}~\bibnamefont{Maselli}},
  \bibinfo{author}{\bibfnamefont{P.}~\bibnamefont{Pani}},
  \bibinfo{author}{\bibfnamefont{V.}~\bibnamefont{Cardoso}},
  \bibinfo{author}{\bibfnamefont{T.}~\bibnamefont{Abdelsalhin}},
  \bibinfo{author}{\bibfnamefont{L.}~\bibnamefont{Gualtieri}},
  \bibnamefont{and} \bibinfo{author}{\bibfnamefont{V.}~\bibnamefont{Ferrari}},
  \bibinfo{journal}{Phys. Rev. Lett.} \textbf{\bibinfo{volume}{120}},
  \bibinfo{pages}{081101} (\bibinfo{year}{2018}), \eprint{1703.10612}.

\bibitem[{\citenamefont{Feng et~al.}(2023)\citenamefont{Feng, Chakraborty, and
  Cardoso}}]{Feng:2022evy}
\bibinfo{author}{\bibfnamefont{J.~C.} \bibnamefont{Feng}},
  \bibinfo{author}{\bibfnamefont{S.}~\bibnamefont{Chakraborty}},
  \bibnamefont{and} \bibinfo{author}{\bibfnamefont{V.}~\bibnamefont{Cardoso}},
  \bibinfo{journal}{Phys. Rev. D} \textbf{\bibinfo{volume}{107}},
  \bibinfo{pages}{044050} (\bibinfo{year}{2023}), \eprint{2211.05261}.

\bibitem[{\citenamefont{Pina et~al.}(2022)\citenamefont{Pina, Orselli, and
  Pica}}]{Pina:2022dye}
\bibinfo{author}{\bibfnamefont{D.~M.} \bibnamefont{Pina}},
  \bibinfo{author}{\bibfnamefont{M.}~\bibnamefont{Orselli}}, \bibnamefont{and}
  \bibinfo{author}{\bibfnamefont{D.}~\bibnamefont{Pica}},
  \bibinfo{journal}{Phys. Rev. D} \textbf{\bibinfo{volume}{106}},
  \bibinfo{pages}{084012} (\bibinfo{year}{2022}), \eprint{2204.08841}.

\bibitem[{\citenamefont{Bozzola and Paschalidis}(2021)}]{Bozzola:2020mjx}
\bibinfo{author}{\bibfnamefont{G.}~\bibnamefont{Bozzola}} \bibnamefont{and}
  \bibinfo{author}{\bibfnamefont{V.}~\bibnamefont{Paschalidis}},
  \bibinfo{journal}{Phys. Rev. Lett.} \textbf{\bibinfo{volume}{126}},
  \bibinfo{pages}{041103} (\bibinfo{year}{2021}), \eprint{2006.15764}.

\bibitem[{\citenamefont{Neves}(2020)}]{Neves2020M87}
\bibinfo{author}{\bibfnamefont{J.~C.~S.} \bibnamefont{Neves}},
  \bibinfo{journal}{Eur. Phys. J. C} \textbf{\bibinfo{volume}{80}},
  \bibinfo{pages}{717} (\bibinfo{year}{2020}), \eprint{2005.00483}.

\bibitem[{\citenamefont{Zaja\v{c}ek et~al.}(2019)\citenamefont{Zaja\v{c}ek,
  Tursunov, Eckart, Britzen, Hackmann, Karas, Stuchl\'\i{}k, Czerny, and
  Zensus}}]{Zajacek2018SgrAReview}
\bibinfo{author}{\bibfnamefont{M.}~\bibnamefont{Zaja\v{c}ek}},
  \bibinfo{author}{\bibfnamefont{A.}~\bibnamefont{Tursunov}},
  \bibinfo{author}{\bibfnamefont{A.}~\bibnamefont{Eckart}},
  \bibinfo{author}{\bibfnamefont{S.}~\bibnamefont{Britzen}},
  \bibinfo{author}{\bibfnamefont{E.}~\bibnamefont{Hackmann}},
  \bibinfo{author}{\bibfnamefont{V.}~\bibnamefont{Karas}},
  \bibinfo{author}{\bibfnamefont{Z.}~\bibnamefont{Stuchl\'\i{}k}},
  \bibinfo{author}{\bibfnamefont{B.}~\bibnamefont{Czerny}}, \bibnamefont{and}
  \bibinfo{author}{\bibfnamefont{J.~A.} \bibnamefont{Zensus}},
  \bibinfo{journal}{J. Phys. Conf. Ser.} \textbf{\bibinfo{volume}{1258}},
  \bibinfo{pages}{012031} (\bibinfo{year}{2019}), \eprint{1812.03574}.

\bibitem[{\citenamefont{Zakharov}(2022)}]{Zakharov2021EHT2017}
\bibinfo{author}{\bibfnamefont{A.~F.} \bibnamefont{Zakharov}},
  \bibinfo{journal}{Universe} \textbf{\bibinfo{volume}{8}},
  \bibinfo{pages}{141} (\bibinfo{year}{2022}), \eprint{2108.01533}.

\bibitem[{\citenamefont{Maselli et~al.}(2015)\citenamefont{Maselli, Gualtieri,
  Pani, Stella, and Ferrari}}]{Maselli:2014fca}
\bibinfo{author}{\bibfnamefont{A.}~\bibnamefont{Maselli}},
  \bibinfo{author}{\bibfnamefont{L.}~\bibnamefont{Gualtieri}},
  \bibinfo{author}{\bibfnamefont{P.}~\bibnamefont{Pani}},
  \bibinfo{author}{\bibfnamefont{L.}~\bibnamefont{Stella}}, \bibnamefont{and}
  \bibinfo{author}{\bibfnamefont{V.}~\bibnamefont{Ferrari}},
  \bibinfo{journal}{Astrophys. J.} \textbf{\bibinfo{volume}{801}},
  \bibinfo{pages}{115} (\bibinfo{year}{2015}), \eprint{1412.3473}.

\bibitem[{\citenamefont{Stuchl\'\i{}k and Kotrlov\'a}(2009)}]{Stuchlik:2008fy}
\bibinfo{author}{\bibfnamefont{Z.}~\bibnamefont{Stuchl\'\i{}k}}
  \bibnamefont{and}
  \bibinfo{author}{\bibfnamefont{A.}~\bibnamefont{Kotrlov\'a}},
  \bibinfo{journal}{Gen. Rel. Grav.} \textbf{\bibinfo{volume}{41}},
  \bibinfo{pages}{1305} (\bibinfo{year}{2009}), \eprint{0812.5066}.

\bibitem[{\citenamefont{Banerjee et~al.}(2017)\citenamefont{Banerjee,
  Chakraborty, and SenGupta}}]{Banerjee:2017hzw}
\bibinfo{author}{\bibfnamefont{I.}~\bibnamefont{Banerjee}},
  \bibinfo{author}{\bibfnamefont{S.}~\bibnamefont{Chakraborty}},
  \bibnamefont{and} \bibinfo{author}{\bibfnamefont{S.}~\bibnamefont{SenGupta}},
  \bibinfo{journal}{Phys. Rev. D} \textbf{\bibinfo{volume}{96}},
  \bibinfo{pages}{084035} (\bibinfo{year}{2017}), \eprint{1707.04494}.

\bibitem[{\citenamefont{Banerjee et~al.}(2021)\citenamefont{Banerjee,
  Chakraborty, and SenGupta}}]{Banerjee:2021aln}
\bibinfo{author}{\bibfnamefont{I.}~\bibnamefont{Banerjee}},
  \bibinfo{author}{\bibfnamefont{S.}~\bibnamefont{Chakraborty}},
  \bibnamefont{and} \bibinfo{author}{\bibfnamefont{S.}~\bibnamefont{SenGupta}},
  \bibinfo{journal}{JCAP} \textbf{\bibinfo{volume}{09}}, \bibinfo{pages}{037}
  (\bibinfo{year}{2021}), \eprint{2105.06636}.

\bibitem[{\citenamefont{Barausse and Yagi}(2015)}]{Barausse2015wia}
\bibinfo{author}{\bibfnamefont{E.}~\bibnamefont{Barausse}} \bibnamefont{and}
  \bibinfo{author}{\bibfnamefont{K.}~\bibnamefont{Yagi}},
  \bibinfo{journal}{Phys. Rev. Lett.} \textbf{\bibinfo{volume}{115}},
  \bibinfo{pages}{211105} (\bibinfo{year}{2015}), \eprint{1509.04539}.

\bibitem[{\citenamefont{Andriot and Lucena~G\'omez}(2017)}]{Andriot2017}
\bibinfo{author}{\bibfnamefont{D.}~\bibnamefont{Andriot}} \bibnamefont{and}
  \bibinfo{author}{\bibfnamefont{G.}~\bibnamefont{Lucena~G\'omez}},
  \bibinfo{journal}{JCAP} \textbf{\bibinfo{volume}{06}}, \bibinfo{pages}{048}
  (\bibinfo{year}{2017}), \bibinfo{note}{[Erratum: JCAP 05, E01 (2019)]},
  \eprint{1704.07392}.

\bibitem[{\citenamefont{Chakraborty et~al.}(2018)\citenamefont{Chakraborty,
  Chakravarti, Bose, and SenGupta}}]{Chakraborty2017}
\bibinfo{author}{\bibfnamefont{S.}~\bibnamefont{Chakraborty}},
  \bibinfo{author}{\bibfnamefont{K.}~\bibnamefont{Chakravarti}},
  \bibinfo{author}{\bibfnamefont{S.}~\bibnamefont{Bose}}, \bibnamefont{and}
  \bibinfo{author}{\bibfnamefont{S.}~\bibnamefont{SenGupta}},
  \bibinfo{journal}{Phys. Rev. D} \textbf{\bibinfo{volume}{97}},
  \bibinfo{pages}{104053} (\bibinfo{year}{2018}), \eprint{1710.05188}.

\bibitem[{\citenamefont{Chakravarti et~al.}(2020)\citenamefont{Chakravarti,
  Chakraborty, Phukon, Bose, and SenGupta}}]{Chakravarti2019}
\bibinfo{author}{\bibfnamefont{K.}~\bibnamefont{Chakravarti}},
  \bibinfo{author}{\bibfnamefont{S.}~\bibnamefont{Chakraborty}},
  \bibinfo{author}{\bibfnamefont{K.~S.} \bibnamefont{Phukon}},
  \bibinfo{author}{\bibfnamefont{S.}~\bibnamefont{Bose}}, \bibnamefont{and}
  \bibinfo{author}{\bibfnamefont{S.}~\bibnamefont{SenGupta}},
  \bibinfo{journal}{Class. Quant. Grav.} \textbf{\bibinfo{volume}{37}},
  \bibinfo{pages}{105004} (\bibinfo{year}{2020}), \eprint{1903.10159}.

\bibitem[{\citenamefont{Gu et~al.}(2024)\citenamefont{Gu, Wang, and
  Shao}}]{Gu:2023eaa}
\bibinfo{author}{\bibfnamefont{H.-P.} \bibnamefont{Gu}},
  \bibinfo{author}{\bibfnamefont{H.-T.} \bibnamefont{Wang}}, \bibnamefont{and}
  \bibinfo{author}{\bibfnamefont{L.}~\bibnamefont{Shao}},
  \bibinfo{journal}{Phys. Rev. D} \textbf{\bibinfo{volume}{109}},
  \bibinfo{pages}{024058} (\bibinfo{year}{2024}), \eprint{2310.10447}.

\bibitem[{\citenamefont{Gupta et~al.}(2021)\citenamefont{Gupta, Spieksma, Pang,
  Koekoek, and Broeck}}]{Gupta:2021rod}
\bibinfo{author}{\bibfnamefont{P.~K.} \bibnamefont{Gupta}},
  \bibinfo{author}{\bibfnamefont{T.~F.~M.} \bibnamefont{Spieksma}},
  \bibinfo{author}{\bibfnamefont{P.~T.~H.} \bibnamefont{Pang}},
  \bibinfo{author}{\bibfnamefont{G.}~\bibnamefont{Koekoek}}, \bibnamefont{and}
  \bibinfo{author}{\bibfnamefont{C.~V.~D.} \bibnamefont{Broeck}},
  \bibinfo{journal}{Phys. Rev. D} \textbf{\bibinfo{volume}{104}},
  \bibinfo{pages}{063041} (\bibinfo{year}{2021}), \eprint{2107.12111}.

\bibitem[{\citenamefont{Carullo et~al.}(2022)\citenamefont{Carullo, Laghi,
  Johnson-McDaniel, Del~Pozzo, Dias, Godazgar, and Santos}}]{Carullo:2021oxn}
\bibinfo{author}{\bibfnamefont{G.}~\bibnamefont{Carullo}},
  \bibinfo{author}{\bibfnamefont{D.}~\bibnamefont{Laghi}},
  \bibinfo{author}{\bibfnamefont{N.~K.} \bibnamefont{Johnson-McDaniel}},
  \bibinfo{author}{\bibfnamefont{W.}~\bibnamefont{Del~Pozzo}},
  \bibinfo{author}{\bibfnamefont{O.~J.~C.} \bibnamefont{Dias}},
  \bibinfo{author}{\bibfnamefont{M.}~\bibnamefont{Godazgar}}, \bibnamefont{and}
  \bibinfo{author}{\bibfnamefont{J.~E.} \bibnamefont{Santos}},
  \bibinfo{journal}{Phys. Rev. D} \textbf{\bibinfo{volume}{105}},
  \bibinfo{pages}{062009} (\bibinfo{year}{2022}), \eprint{2109.13961}.

\bibitem[{\citenamefont{Mishra et~al.}(2022)\citenamefont{Mishra, Ghosh, and
  Chakraborty}}]{MishraAkash_I}
\bibinfo{author}{\bibfnamefont{A.~K.} \bibnamefont{Mishra}},
  \bibinfo{author}{\bibfnamefont{A.}~\bibnamefont{Ghosh}}, \bibnamefont{and}
  \bibinfo{author}{\bibfnamefont{S.}~\bibnamefont{Chakraborty}},
  \bibinfo{journal}{Eur. Phys. J. C} \textbf{\bibinfo{volume}{82}},
  \bibinfo{pages}{820} (\bibinfo{year}{2022}), \eprint{2106.05558}.

\bibitem[{\citenamefont{Mishra et~al.}(2024)\citenamefont{Mishra, Carullo, and
  Chakraborty}}]{MishraAkash_II}
\bibinfo{author}{\bibfnamefont{A.~K.} \bibnamefont{Mishra}},
  \bibinfo{author}{\bibfnamefont{G.}~\bibnamefont{Carullo}}, \bibnamefont{and}
  \bibinfo{author}{\bibfnamefont{S.}~\bibnamefont{Chakraborty}},
  \bibinfo{journal}{Phys. Rev. D} \textbf{\bibinfo{volume}{109}},
  \bibinfo{pages}{024025} (\bibinfo{year}{2024}), \eprint{2311.03556}.

\bibitem[{\citenamefont{Deka et~al.}(2024)\citenamefont{Deka, Chakraborty,
  Kapadia, Shaikh, and Ajith}}]{UddeeptaDeka2024}
\bibinfo{author}{\bibfnamefont{U.}~\bibnamefont{Deka}},
  \bibinfo{author}{\bibfnamefont{S.}~\bibnamefont{Chakraborty}},
  \bibinfo{author}{\bibfnamefont{S.~J.} \bibnamefont{Kapadia}},
  \bibinfo{author}{\bibfnamefont{M.~A.} \bibnamefont{Shaikh}},
  \bibnamefont{and} \bibinfo{author}{\bibfnamefont{P.}~\bibnamefont{Ajith}}
  (\bibinfo{year}{2024}), \eprint{2401.06553}.

\bibitem[{\citenamefont{Dey et~al.}(2020)\citenamefont{Dey, Chakraborty, and
  Afshordi}}]{Dey:2020lhq}
\bibinfo{author}{\bibfnamefont{R.}~\bibnamefont{Dey}},
  \bibinfo{author}{\bibfnamefont{S.}~\bibnamefont{Chakraborty}},
  \bibnamefont{and} \bibinfo{author}{\bibfnamefont{N.}~\bibnamefont{Afshordi}},
  \bibinfo{journal}{Phys. Rev. D} \textbf{\bibinfo{volume}{101}},
  \bibinfo{pages}{104014} (\bibinfo{year}{2020}), \eprint{2001.01301}.

\bibitem[{\citenamefont{Dey et~al.}(2021)\citenamefont{Dey, Biswas, and
  Chakraborty}}]{Dey:2020pth}
\bibinfo{author}{\bibfnamefont{R.}~\bibnamefont{Dey}},
  \bibinfo{author}{\bibfnamefont{S.}~\bibnamefont{Biswas}}, \bibnamefont{and}
  \bibinfo{author}{\bibfnamefont{S.}~\bibnamefont{Chakraborty}},
  \bibinfo{journal}{Phys. Rev. D} \textbf{\bibinfo{volume}{103}},
  \bibinfo{pages}{084019} (\bibinfo{year}{2021}), \eprint{2010.07966}.

\bibitem[{\citenamefont{Abbott et~al.}(2018)}]{KAGRA:2013rdx}
\bibinfo{author}{\bibfnamefont{B.~P.} \bibnamefont{Abbott}}
  \bibnamefont{et~al.} (\bibinfo{collaboration}{KAGRA, LIGO Scientific, Virgo,
  VIRGO}), \bibinfo{journal}{Living Rev. Rel.} \textbf{\bibinfo{volume}{21}},
  \bibinfo{pages}{3} (\bibinfo{year}{2018}), \eprint{1304.0670}.

\bibitem[{\citenamefont{Teukolsky}(1972)}]{Teukolsky1972}
\bibinfo{author}{\bibfnamefont{S.~A.} \bibnamefont{Teukolsky}},
  \bibinfo{journal}{Phys. Rev. Lett.} \textbf{\bibinfo{volume}{29}},
  \bibinfo{pages}{1114} (\bibinfo{year}{1972}),
  \urlprefix\url{https://link.aps.org/doi/10.1103/PhysRevLett.29.1114}.

\bibitem[{\citenamefont{Chakraborty et~al.}(2021)\citenamefont{Chakraborty,
  Datta, and Sau}}]{Chakraborty:2021gdf}
\bibinfo{author}{\bibfnamefont{S.}~\bibnamefont{Chakraborty}},
  \bibinfo{author}{\bibfnamefont{S.}~\bibnamefont{Datta}}, \bibnamefont{and}
  \bibinfo{author}{\bibfnamefont{S.}~\bibnamefont{Sau}},
  \bibinfo{journal}{Phys. Rev. D} \textbf{\bibinfo{volume}{104}},
  \bibinfo{pages}{104001} (\bibinfo{year}{2021}), \eprint{2103.12430}.

\bibitem[{\citenamefont{Chamblin et~al.}(2001)\citenamefont{Chamblin, Reall,
  Shinkai, and Shiromizu}}]{Chamblin:2000ra}
\bibinfo{author}{\bibfnamefont{A.}~\bibnamefont{Chamblin}},
  \bibinfo{author}{\bibfnamefont{H.~S.} \bibnamefont{Reall}},
  \bibinfo{author}{\bibfnamefont{H.-a.} \bibnamefont{Shinkai}},
  \bibnamefont{and}
  \bibinfo{author}{\bibfnamefont{T.}~\bibnamefont{Shiromizu}},
  \bibinfo{journal}{Phys. Rev. D} \textbf{\bibinfo{volume}{63}},
  \bibinfo{pages}{064015} (\bibinfo{year}{2001}), \eprint{hep-th/0008177}.

\bibitem[{\citenamefont{Banerjee et~al.}(2022)\citenamefont{Banerjee,
  Chakraborty, and SenGupta}}]{Banerjee:2022jog}
\bibinfo{author}{\bibfnamefont{I.}~\bibnamefont{Banerjee}},
  \bibinfo{author}{\bibfnamefont{S.}~\bibnamefont{Chakraborty}},
  \bibnamefont{and} \bibinfo{author}{\bibfnamefont{S.}~\bibnamefont{SenGupta}},
  \bibinfo{journal}{Phys. Rev. D} \textbf{\bibinfo{volume}{106}},
  \bibinfo{pages}{084051} (\bibinfo{year}{2022}), \eprint{2207.09003}.

\bibitem[{\citenamefont{Banerjee et~al.}(2020)\citenamefont{Banerjee,
  Chakraborty, and SenGupta}}]{Banerjee:2019nnj}
\bibinfo{author}{\bibfnamefont{I.}~\bibnamefont{Banerjee}},
  \bibinfo{author}{\bibfnamefont{S.}~\bibnamefont{Chakraborty}},
  \bibnamefont{and} \bibinfo{author}{\bibfnamefont{S.}~\bibnamefont{SenGupta}},
  \bibinfo{journal}{Phys. Rev. D} \textbf{\bibinfo{volume}{101}},
  \bibinfo{pages}{041301} (\bibinfo{year}{2020}), \eprint{1909.09385}.

\bibitem[{\citenamefont{Arun et~al.}(2009)\citenamefont{Arun, Buonanno, Faye,
  and Ochsner}}]{ABFO08}
\bibinfo{author}{\bibfnamefont{K.~G.} \bibnamefont{Arun}},
  \bibinfo{author}{\bibfnamefont{A.}~\bibnamefont{Buonanno}},
  \bibinfo{author}{\bibfnamefont{G.}~\bibnamefont{Faye}}, \bibnamefont{and}
  \bibinfo{author}{\bibfnamefont{E.}~\bibnamefont{Ochsner}},
  \bibinfo{journal}{Phys. Rev. D} \textbf{\bibinfo{volume}{79}},
  \bibinfo{pages}{104023} (\bibinfo{year}{2009}), \eprint{0810.5336}.

\bibitem[{\citenamefont{Buonanno et~al.}(2009)\citenamefont{Buonanno, Iyer,
  Ochsner, Pan, and Sathyaprakash}}]{BIOPS2009}
\bibinfo{author}{\bibfnamefont{A.}~\bibnamefont{Buonanno}},
  \bibinfo{author}{\bibfnamefont{B.}~\bibnamefont{Iyer}},
  \bibinfo{author}{\bibfnamefont{E.}~\bibnamefont{Ochsner}},
  \bibinfo{author}{\bibfnamefont{Y.}~\bibnamefont{Pan}}, \bibnamefont{and}
  \bibinfo{author}{\bibfnamefont{B.~S.} \bibnamefont{Sathyaprakash}},
  \bibinfo{journal}{Phys. Rev. D} \textbf{\bibinfo{volume}{80}},
  \bibinfo{pages}{084043} (\bibinfo{year}{2009}), \eprint{0907.0700}.

\bibitem[{\citenamefont{Mishra et~al.}(2016)\citenamefont{Mishra, Kela, Arun,
  and Faye}}]{Mishra:2016whh}
\bibinfo{author}{\bibfnamefont{C.~K.} \bibnamefont{Mishra}},
  \bibinfo{author}{\bibfnamefont{A.}~\bibnamefont{Kela}},
  \bibinfo{author}{\bibfnamefont{K.~G.} \bibnamefont{Arun}}, \bibnamefont{and}
  \bibinfo{author}{\bibfnamefont{G.}~\bibnamefont{Faye}},
  \bibinfo{journal}{Phys. Rev. D} \textbf{\bibinfo{volume}{93}},
  \bibinfo{pages}{084054} (\bibinfo{year}{2016}), \eprint{1601.05588}.

\bibitem[{\citenamefont{Mukherjee et~al.}(2022)\citenamefont{Mukherjee, Datta,
  Tiwari, Phukon, and Bose}}]{Samanwaya2022}
\bibinfo{author}{\bibfnamefont{S.}~\bibnamefont{Mukherjee}},
  \bibinfo{author}{\bibfnamefont{S.}~\bibnamefont{Datta}},
  \bibinfo{author}{\bibfnamefont{S.}~\bibnamefont{Tiwari}},
  \bibinfo{author}{\bibfnamefont{K.~S.} \bibnamefont{Phukon}},
  \bibnamefont{and} \bibinfo{author}{\bibfnamefont{S.}~\bibnamefont{Bose}},
  \bibinfo{journal}{Phys. Rev. D} \textbf{\bibinfo{volume}{106}},
  \bibinfo{pages}{104032} (\bibinfo{year}{2022}), \eprint{2202.08661}.

\bibitem[{\citenamefont{Chakravarti et~al.}(2019)\citenamefont{Chakravarti,
  Chakraborty, Bose, and SenGupta}}]{Chakravarti:2018vlt}
\bibinfo{author}{\bibfnamefont{K.}~\bibnamefont{Chakravarti}},
  \bibinfo{author}{\bibfnamefont{S.}~\bibnamefont{Chakraborty}},
  \bibinfo{author}{\bibfnamefont{S.}~\bibnamefont{Bose}}, \bibnamefont{and}
  \bibinfo{author}{\bibfnamefont{S.}~\bibnamefont{SenGupta}},
  \bibinfo{journal}{Phys. Rev. D} \textbf{\bibinfo{volume}{99}},
  \bibinfo{pages}{024036} (\bibinfo{year}{2019}), \eprint{1811.11364}.

\bibitem[{\citenamefont{{LIGO Scientific Collaboration}}(2018)}]{lalsuite}
\bibinfo{author}{\bibnamefont{{LIGO Scientific Collaboration}}},
  \emph{\bibinfo{title}{{LIGO} {A}lgorithm {L}ibrary - {LALS}uite}},
  \bibinfo{howpublished}{free software (GPL)} (\bibinfo{year}{2018}).

\bibitem[{\citenamefont{Ashton et~al.}(2019)}]{Ashton:2018jfp}
\bibinfo{author}{\bibfnamefont{G.}~\bibnamefont{Ashton}} \bibnamefont{et~al.},
  \bibinfo{journal}{Astrophys. J. Suppl.} \textbf{\bibinfo{volume}{241}},
  \bibinfo{pages}{27} (\bibinfo{year}{2019}), \eprint{1811.02042}.

\bibitem[{\citenamefont{Favata et~al.}(2022)\citenamefont{Favata, Kim, Arun,
  Kim, and Lee}}]{Favata:2021vhw}
\bibinfo{author}{\bibfnamefont{M.}~\bibnamefont{Favata}},
  \bibinfo{author}{\bibfnamefont{C.}~\bibnamefont{Kim}},
  \bibinfo{author}{\bibfnamefont{K.~G.} \bibnamefont{Arun}},
  \bibinfo{author}{\bibfnamefont{J.}~\bibnamefont{Kim}}, \bibnamefont{and}
  \bibinfo{author}{\bibfnamefont{H.~W.} \bibnamefont{Lee}},
  \bibinfo{journal}{Phys. Rev. D} \textbf{\bibinfo{volume}{105}},
  \bibinfo{pages}{023003} (\bibinfo{year}{2022}), \eprint{2108.05861}.

\bibitem[{\citenamefont{Abbott et~al.}(2023)}]{GWOSC-3}
\bibinfo{author}{\bibfnamefont{R.}~\bibnamefont{Abbott}} \bibnamefont{et~al.}
  (\bibinfo{collaboration}{KAGRA, VIRGO, LIGO Scientific}),
  \bibinfo{journal}{Astrophys. J. Suppl.} \textbf{\bibinfo{volume}{267}},
  \bibinfo{pages}{29} (\bibinfo{year}{2023}), \eprint{2302.03676}.

\bibitem[{\citenamefont{Abbott et~al.}(2020{\natexlab{c}})}]{GW190412}
\bibinfo{author}{\bibfnamefont{R.}~\bibnamefont{Abbott}} \bibnamefont{et~al.}
  (\bibinfo{collaboration}{LIGO Scientific, Virgo}), \bibinfo{journal}{Phys.
  Rev. D} \textbf{\bibinfo{volume}{102}}, \bibinfo{pages}{043015}
  (\bibinfo{year}{2020}{\natexlab{c}}), \eprint{2004.08342}.

\bibitem[{\citenamefont{Romero-Shaw et~al.}(2020)}]{Romero-Shaw:2020owr}
\bibinfo{author}{\bibfnamefont{I.~M.} \bibnamefont{Romero-Shaw}}
  \bibnamefont{et~al.}, \bibinfo{journal}{Mon. Not. Roy. Astron. Soc.}
  \textbf{\bibinfo{volume}{499}}, \bibinfo{pages}{3295} (\bibinfo{year}{2020}),
  \eprint{arXiv:2006.00714 [astro-ph.IM]}.

\bibitem[{\citenamefont{{Harris} et~al.}(2020)\citenamefont{{Harris},
  {Millman}, {van der Walt}, {Gommers}, {Virtanen}, {Cournapeau}, {Wieser},
  {Taylor}, {Berg}, {Smith} et~al.}}]{2020Natur.585..357H}
\bibinfo{author}{\bibfnamefont{C.~R.} \bibnamefont{{Harris}}},
  \bibinfo{author}{\bibfnamefont{K.~J.} \bibnamefont{{Millman}}},
  \bibinfo{author}{\bibfnamefont{S.~J.} \bibnamefont{{van der Walt}}},
  \bibinfo{author}{\bibfnamefont{R.}~\bibnamefont{{Gommers}}},
  \bibinfo{author}{\bibfnamefont{P.}~\bibnamefont{{Virtanen}}},
  \bibinfo{author}{\bibfnamefont{D.}~\bibnamefont{{Cournapeau}}},
  \bibinfo{author}{\bibfnamefont{E.}~\bibnamefont{{Wieser}}},
  \bibinfo{author}{\bibfnamefont{J.}~\bibnamefont{{Taylor}}},
  \bibinfo{author}{\bibfnamefont{S.}~\bibnamefont{{Berg}}},
  \bibinfo{author}{\bibfnamefont{N.~J.} \bibnamefont{{Smith}}},
  \bibnamefont{et~al.}, \bibinfo{journal}{\nat} \textbf{\bibinfo{volume}{585}},
  \bibinfo{pages}{357} (\bibinfo{year}{2020}), \eprint{arXiv:2006.10256
  [cs.MS]}.

\bibitem[{\citenamefont{Hoy and Raymond}(2021)}]{Hoy:2020vys}
\bibinfo{author}{\bibfnamefont{C.}~\bibnamefont{Hoy}} \bibnamefont{and}
  \bibinfo{author}{\bibfnamefont{V.}~\bibnamefont{Raymond}},
  \bibinfo{journal}{SoftwareX} \textbf{\bibinfo{volume}{15}},
  \bibinfo{pages}{100765} (\bibinfo{year}{2021}), \eprint{arXiv:2006.06639
  [astro-ph.IM]}.

\bibitem[{\citenamefont{{Hunter}}(2007)}]{2007CSE.....9...90H}
\bibinfo{author}{\bibfnamefont{J.~D.} \bibnamefont{{Hunter}}},
  \bibinfo{journal}{Computing in Science and Engineering}
  \textbf{\bibinfo{volume}{9}}, \bibinfo{pages}{90} (\bibinfo{year}{2007}).

\bibitem[{\citenamefont{Kluyver et~al.}(2016)\citenamefont{Kluyver,
  Ragan-Kelley, P{\'e}rez, Granger, Bussonnier, Frederic, Kelley, Hamrick,
  Grout, Corlay et~al.}}]{soton403913}
\bibinfo{author}{\bibfnamefont{T.}~\bibnamefont{Kluyver}},
  \bibinfo{author}{\bibfnamefont{B.}~\bibnamefont{Ragan-Kelley}},
  \bibinfo{author}{\bibfnamefont{F.}~\bibnamefont{P{\'e}rez}},
  \bibinfo{author}{\bibfnamefont{B.}~\bibnamefont{Granger}},
  \bibinfo{author}{\bibfnamefont{M.}~\bibnamefont{Bussonnier}},
  \bibinfo{author}{\bibfnamefont{J.}~\bibnamefont{Frederic}},
  \bibinfo{author}{\bibfnamefont{K.}~\bibnamefont{Kelley}},
  \bibinfo{author}{\bibfnamefont{J.}~\bibnamefont{Hamrick}},
  \bibinfo{author}{\bibfnamefont{J.}~\bibnamefont{Grout}},
  \bibinfo{author}{\bibfnamefont{S.}~\bibnamefont{Corlay}},
  \bibnamefont{et~al.}, in \emph{\bibinfo{booktitle}{Positioning and Power in
  Academic Publishing: Players, Agents and Agendas}}, edited by
  \bibinfo{editor}{\bibfnamefont{F.}~\bibnamefont{Loizides}} \bibnamefont{and}
  \bibinfo{editor}{\bibfnamefont{B.}~\bibnamefont{Scmidt}}
  (\bibinfo{publisher}{IOS Press}, \bibinfo{year}{2016}), pp.
  \bibinfo{pages}{87--90}.

\bibitem[{\citenamefont{Speagle}(2020)}]{speagle2020dynesty}
\bibinfo{author}{\bibfnamefont{J.~S.} \bibnamefont{Speagle}},
  \bibinfo{journal}{Monthly Notices of the Royal Astronomical Society}
  \textbf{\bibinfo{volume}{493}}, \bibinfo{pages}{3132} (\bibinfo{year}{2020}).

\bibitem[{\citenamefont{Tichy et~al.}(2000)\citenamefont{Tichy, Flanagan, and
  Poisson}}]{Tichy:1999pv}
\bibinfo{author}{\bibfnamefont{W.}~\bibnamefont{Tichy}},
  \bibinfo{author}{\bibfnamefont{E.~E.} \bibnamefont{Flanagan}},
  \bibnamefont{and} \bibinfo{author}{\bibfnamefont{E.}~\bibnamefont{Poisson}},
  \bibinfo{journal}{Phys. Rev. D} \textbf{\bibinfo{volume}{61}},
  \bibinfo{pages}{104015} (\bibinfo{year}{2000}), \eprint{gr-qc/9912075}.

\bibitem[{\citenamefont{Damour et~al.}(1998)\citenamefont{Damour, Iyer, and
  Sathyaprakash}}]{DIS1997}
\bibinfo{author}{\bibfnamefont{T.}~\bibnamefont{Damour}},
  \bibinfo{author}{\bibfnamefont{B.~R.} \bibnamefont{Iyer}}, \bibnamefont{and}
  \bibinfo{author}{\bibfnamefont{B.~S.} \bibnamefont{Sathyaprakash}},
  \bibinfo{journal}{Phys. Rev. D} \textbf{\bibinfo{volume}{57}},
  \bibinfo{pages}{885} (\bibinfo{year}{1998}), \eprint{gr-qc/9708034}.

\end{thebibliography}
\end{document}